\documentclass[useAMS, a4paper]{mn2e}
\usepackage{mnras_cite}
\usepackage{epsfig, amsmath}
\def\gsim{\ga}

\def\H2{{\rm H}$_2$}

\newcommand{\data}{d}
\newcommand{\pdf}{P}

\newcommand{\params}{\theta}
\newcommand{\fdil}{f_{\rm dil}}
\newcommand{\logM}{\log M_{\rm min}}
\newcommand{\Mmin}{M_{\rm min}}
\newcommand{\Mgas}{M_{\rm gas}}
\newcommand{\Mstars}{M_\star}
\newcommand{\Mstruc}{M_{\rm struct}}

\newcommand{\half}{0.50}

\def\simlt{\lower.5ex\hbox{$\; \buildrel < \over \sim \;$}}
\def\simgt{\lower.5ex\hbox{$\; \buildrel > \over \sim \;$}}
\def\simpropto{\lower.2ex\hbox{$\; \buildrel \propto \over \sim \;$}}

\begin{document}
\title[Monolithic or hierarchical star formation]{Monolithic or hierarchical star formation? A new statistical analysis}
\author[Kampakoglou, Trotta and Silk]{Marios Kampakoglou\thanks{E-mail address: {\tt mariosk@astro.ox.ac.uk}},
Roberto Trotta\thanks{E-mail address: {\tt rxt@astro.ox.ac.uk}}
and Joseph Silk\thanks{E-mail address: {\tt silk@astro.ox.ac.uk}}}
\maketitle

\begin{abstract}
We consider an analytic model of cosmic star formation which
incorporates supernova feedback, gas accretion and enriched
outflows, reproducing the history of cosmic star formation,
metallicity, supernovae type II rates and the fraction of baryons
allocated to structures. We present a new statistical treatment of
the available observational data on the star formation rate and
metallicity that accounts for the presence of  possible
systematics. We then employ a Bayesian Markov Chain Monte Carlo
method to compare the predictions of our model with observations
and derive constraints on the 7 free parameters of the model. We
find that the dust correction scheme one chooses to adopt for the
star formation data is critical in determining which scenario is
favoured between  a hierarchical star formation model, where star
formation is prolonged by accretion, infall and merging, and a
monolithic scenario, where star formation is rapid and efficient.
We distinguish between these modes by  defining a characteristic
minimum mass, $M \gsim 10^{11} M_\odot$, in our fiducial model,
for early type galaxies  where star formation occurs efficiently.
Our results indicate that the hierarchical star formation model
can achieve better agreement with the data, but that this requires
a high efficiency of supernova--driven outflows. In a monolithic
model, our analysis points to the need for a mechanism that drives
metal--poor winds, perhaps in the form of supermassive black
hole--induced outflows. Furthermore, the relative absence of star
formation beyond $z \sim 5$ in the monolithic scenario requires an
alternative mechanism to dwarf galaxies for reionizing the
universe at $z \sim 11$, as required by observations of the
microwave background. While the monolithic scenario is less
favoured in terms of its quality--of--fit, it cannot yet be
excluded.
\end{abstract}
\section{Introduction}

Massive galactic spheroids form either by hierarchical build--up
or monolithically. The former scenario is favoured by the theory
of, and evidence for, cold dark matter (CDM), the latter by {\it
some} observations, most notably the evidence for down--sizing in
stellar mass and on chemical evolution time--scales from the
$[\alpha/\text{Fe}]$ enhancement at high spheroid masses
(\pcite{worthey92}). The conventional view of hierarchical
build--up  fits in with semi--analytical galaxy formation
simulations provided that suitable feedback models are prescribed
(\pcite{croton06}, \pcite{bower06}). Gas cooling drives star
formation below a critical halo mass and AGN quenching of star
formation occurs at higher masses where the cooling is
inefficient. This model reproduces the shape of the galaxy
luminosity function at  the high mass end, and also produces old,
red massive galaxies. Not yet addressed in these studies is the
issue of whether these massive galaxies can be formed sufficiently
rapidly,  the resolution of which  may lend support to  a
monolithic formation scenario. Major mergers are a plausible
ingredient of a monolithic model, for triggering rapid star
formation.

Not all studies agree on the role of major mergers.  Not all
luminous starbursts are triggered by such mergers. Indeed, in the
case of  disk galaxy formation, their role is likely to be small.
There is evidence that major mergers
are not exclusively
responsible for the dominant  star formation episodes in
starbursts forming stars at up to 200 M$_\odot$/yr where thick
disks are seen (e.g. \pcite{hammer05}). Such high star formation
rates, also found  in disks at $z\sim 2$ by \pcite{forster06}, may
favour an interpretation in terms of bulge formation since the
disks are already present. The observed major merger rate is small out to
$z\sim 1.2$ despite the strong increase in the comoving star
formation rate \cite{lotz06}. Of course, ULIRGs, with star
formation rates of $\simgt 500 \rm M_\odot$/yr, are  examples of major
merger--induced star formation. However, for the bulk of star
formation in the early universe, there is little evidence to
suggest that major mergers play a significant role.  This motivates a hierarchical picture of prolonged minor mergers to build up mass.

One can  ask more generally whether minor mergers or gas cloud accretion are
responsible for the gas supply. With regard to minor mergers, the
answer seems to be negative, because the time--scale for supplying
the gas would be many dynamical time--scales. Yet there are two
persuasive arguments for a short star formation time--scale in
massive galaxies. SED modelling suggests down--sizing, the most
massive galaxies forming first and with star formation
time--scales of order a dynamical time. The $\rm [{\alpha}/Fe]$
enhancement with increasing spheroid mass independently reinforces
this conclusion, since most of the star formation must have
occurred before dilution of the star--forming gas by supernovae
type Ia ejecta occurred. This result is not, or at least not yet,
found in  semi--analytical modelling of massive galaxy formation.
It provides a strong argument for monolithic formation of massive
galaxies, although one cannot of course exclude the possibility
that hierarchical formation models will  reproduce a similar
result once more complex star formation rules are introduced.
Indeed,  simple spherically symmetric accretion models have been
proposed (cf. \pcite{birnboim07}) that allow the gas to accumulate
in an isothermal halo atmosphere prior to an accretion--triggered
burst of star formation. In essence, this approach introduces a
monolothic formation model in combination with hierarchical gas
accumulation. Unfortunately there is little evidence today for
such gas--rich halos, suggesting that this process, if important,
would only have been influential at very early epochs.

To save hierarchical models, one needs an efficient way of
converting gas into stars. The typical gas--to--star  e--folding
time in nearby well--studied sites of star formation, namely
spiral galaxies, is several Gyr. By contrast, at a redshift of
$2-3$, the star formation time--scale is as short as 0.2 Gyr from
both spectrophotometric SED filling \cite{mar06} and
$\rm[\alpha/Fe]$ analyses \cite{tho05}. The AGN phenomenon is
invoked for stopping star formation by quenching the gas supply.
This may keep massive galaxies red, but may not be enough to
produce massive galaxies sufficiently early. One resolution may be
to invoke positive feedback from AGN--driven outflows that
overpressure protogalactic clouds and trigger star formation on a
rapid time--scale \cite{silk05}. Whether or not this particular
solution turns out to resolve the dilemma (if indeed the problem
persists in the perspective of improved star formation modelling)
is not the issue we focus on in this paper. Rather we revisit the
case for hierarchical versus monolithic galaxy formation in terms
of accounting for the data on star formation rate and chemical
evolution.

In this paper we apply a novel statistical approach that
rigorously treats the key parameters in semi--analytical galaxy
formation theory, with emphasis on reproducing the cosmic star
formation and chemical evolution histories.  Numerical modelling
via semi--analytical techniques of a large box of the universe
takes up so much computer time and memory that it is impossible to
test the robustness of the results. Here we focus on probing the
key parameter space by means of analytical techniques combined
with appropriate binning of the full data sets. We find that the
characteristic minimum  mass for the building blocks of massive
galaxies plays a central role. Specifically, the dust correction
scheme one chooses to adopt for the cosmic star formation history
data is one of the most critical factors in determining the
balance of evidence in support of  a hierarchical star formation
model as opposed to a monolithic scenario, where star formation
happens predominantly  in massive spheroids. Our results indicate
that the hierarchical star formation model can achieve better
agreement with the data, but that this requires a high efficiency
of supernova--driven outflows. In a monolithic model, our analysis
points to the need for a mechanism that drives metal--poor winds,
perhaps in the form of supermassive black hole--induced outflows.
While the monolithic scenario is less favoured in terms of its
quality--of--fit, it cannot yet be excluded.

This paper is organized as follows: our star formation model is
introduced in section~\ref{sec:model}, and the dependence of the
observable quantities on the free parameters of the model explored
in section~\ref{sec:Sensitivity}. We then describe the data
employed and our statistical procedure in section~\ref{sec:data}.
Our results are presented in section~\ref{sec:results} and our
conclusions discussed in section~\ref{sec:conc}.
Appendix~\ref{app} gives details of our binning procedure for star
formation rate and metallicity data which accounts for undetected
systematics.

\section{Star formation model}
\label{sec:model} In this section we present a physical model of
the cosmic star formation incorporating supernova feedback, gas
accretion and enriched outflows. Our model  builds upon the model
described in \pcite{daigne05}, and more specifically their
``Model~0''. The main difference at the model level is our choice
of using a different chemical evolution model. In fact, we adopt
the instantaneous recycling approximation whereas in
\pcite{daigne05} the metallicity is computed under the delayed
enrichment approximation. There are several differences in terms
of the statistical treatment of the data and the fitting
procedure, that in this work are significantly more sophisticated,
as explained in section~\ref{sec:data} and Appendix~\ref{app}.

\subsection{Governing equations}
The description of baryons in the Universe and the processes that
define the evolution of the baryonic mass are of fundamental
importance for our model. Following \pcite{daigne05}, we employ
three baryon reservoirs in the model, encompassing the
interstellar medium (gas), the mass in stars and the intergalactic
medium (IGM). We denote by $\Mgas$ the mass of gas, by $\Mstars$
the mass in stars and by $\Mstruc = \Mgas + \Mstars$ the total
mass in collapsed structures. The IGM and the structures exchange
mass through accretion and outflow, while the interaction between
stars and gas is governed by star formation and ejection of
enriched gas. In the instantaneous recycling approximation adopted
here, the accretion rate of the mass in stars is simply equal to
the star formation rate (SFR), $\Psi(t)$, i.e.
 \begin{equation}
 \frac{d \Mstars}{dt} = \Psi(t).
 \end{equation}
We then have the following set of differential equations governing
the evolution of the mass in the three reservoirs:
\begin{align}
 \frac{dM_{\rm IGM}}{dt} & = - \frac{d\Mstruc}{dt} \\
 \frac{d\Mstruc}{dt}     & =  a_b(t)- o(t)\\
 \frac{d\Mgas}{dt}       & = {\frac{dM_{\rm struct}}{dt}}-{\frac{dM_{\rm \star}}{dt}}
\end{align}
In the above equations, $a_{\rm b}(t)$ is the rate of baryon
accretion while $o(t)$ is the rate of baryon outflow. The latter
includes two terms, accounting for winds powered by stellar
explosions and supernova ejecta. We neglect supernova ejecta since
this effect was found to be subdominant~\cite{daigne04}. The
relation between physical time $t$ and redshift $z$ is given by
\begin{equation}
\label{eq:ttoz} \frac{dt}{dz}~=~\frac{9.78h^{-1}
\rm{Gyr}}{(1+z)\sqrt{{\Omega}_{\rm \Lambda}+{\Omega}_{\rm
m}(1+z)^{3}}},
\end{equation}
where we have assumed a flat Universe. In the following, we will
use Eq.~\eqref{eq:ttoz} with parameters fixed to the values of the
$\Lambda$CDM concordance model, i.e. a matter density parameter
${\Omega}_{\rm m}~=~0.27$ and a cosmological constant energy
density  ${\Omega}_{\rm \Lambda}~=~0.73$ (both in units of the
critical energy density of the Universe), and we will take for the
Hubble constant $ H_{\rm 0} = 100h~{\rm km/sec/Mpc} = 71$.

\subsection{Accretion}
\label{sec:acc}

We adopt the hierarchical scenario of structure formation, where
small structures are formed first. At redshift $z$, the comoving
density of dark matter halos in the mass range $[M, M+dM]$ is
$f_{\rm ps}(M,z)dM$, normalized in such a way that
\begin{equation}
{\int}_{0}^{\infty}dM M f_{\rm ps}(M,z) = {\rho}_{\rm DM},
\end{equation}
where $\rho_{\rm DM}$ is the comoving dark matter density. The
distribution function of halos $f_{\rm ps}(M,z)$ is computed using
the method described in \pcite{jenkins01} using code provided by
A.~Jenkins. It follows the standard theory \cite{press74}
including the modification of \pcite{sheth99}. We assume a
primordial power spectrum of fluctuation with a power law index
$n_S = 1$ and the fitting formula to the exact transfer function
for non-baryonic cold dark matter given by \pcite{bond84}. For the
rms amplitude we adopt a value $\sigma_{\rm 8} = 0.9$ for mass
density fluctuations in a sphere of radius $8h^{-1}$Mpc.

Using the above expressions for the distribution function of dark
matter halos, we can calculate the fraction of baryons at redshift
$z$ that is allocated to structures, by assuming that the baryon
density is proportional to the dark matter density, with a
proportionality factor given by the ratio of visible to dark
matter density -- in other words, we assume that light traces
matter with no bias. The fraction of baryons in star--forming
structures at redshift $z$ is then given by
\begin{equation}
f_{\rm bar}(z) = \frac{\int_{\Mmin}^{\infty}dMMf_{\rm
ps}(M,z)}{\int_{0}^{\infty}dMMf_{\rm ps}(M,z)}, \label{eq:Bar_frc}
\end{equation}
where $\Mmin$ is a free parameter controlling the minimum mass (in
units of solar masses) of the collapsed structures where star
formation can occur.

The accretion rate is then given by \cite{daigne05}:
\begin{equation}
a_{b}(t) = {\Omega}_{b}\frac{3H_{\rm
0}^{2}}{8{\pi}G}\left(\frac{dt}{dz}\right)^{-1}{\large\vert}{\frac{df_{\rm
bar}}{dz}}{\Large\vert}
\end{equation}
Given a value of $\Mmin$ (that we adopt as a free parameter, see
below), we fix the redshift at which star formation begins,
$z_{\rm init}$ by the requirement that $f_{\rm bar}(z_{\rm init})
= 0.01$. In other words, the first stars form in collapsed haloes
of mass larger than $\Mmin$ when the fraction of baryons allocated
to such structures is more than $1\%$.  We adopt a fixed baryonic
density parameter of $\Omega_{\rm b} = 0.044$ (from the posterior
mean of WMAP 3--years data combined with all other datasets,
\pcite{spergel06}).

\subsection{Outflow}
\label{sec:outflow}
  The adopted stellar initial mass function
(IMF) is of the form
\begin{equation}
\label{eq:x}
 {{\Phi}(m)} = B \left(\frac{m}{M{_\odot}}\right)^{-(1+x)}, \quad \text{for} ~m_{\rm
l}<m<m_{\rm u}
\end{equation}
where the normalization constant $B$ is fixed by the requirement
that
\begin{equation}
\int_{m_l}^{m_u}m{\Phi}(m)dm = \Mstars \label{eq:IMF_norm}
\end{equation}
For the limits of integration we fix $m_{\rm l}=0.1M{_{\odot}}$
and  $m_{\rm u}=100M{_{\odot}}$ \cite{pagel97}. Therefore the only
parameter needed to define the IMF is its power--law index, $x$.
The quantity $x$ is used as a free parameter in this model.

We model the outflow powered by stellar explosions as follows:
\begin{equation}
\label{eq:eps} o(t)~=~\frac{2\epsilon}{v_{\rm esc}^{2}(z)}
\int_{m_0}^{100M_\odot}dm{\Phi}(m){\Psi}(t-{\tau_s}(m))E_{\rm
kin}(m)
\end{equation}
where
 \begin{equation}
 \label{eq:m0} m_0 = {\rm max}(8M_{\rm \odot},m_{\rm d}(t))
\end{equation}
and  ${\Phi}(m)$ is the IMF defined above, ${\tau_s}(m)$ is the
lifetime of a star of mass $m$ and $m_d(t)$ is the mass of stars
that die at age $t$. Furthermore,  $E_{\rm kin}(m)$ is the kinetic
energy released by the explosion of a star of mass $m$, that we
take to be a fixed constant independent of mass, $E_{\rm
kin}(m)~=~10^{51}$ ergs (a mass--dependence could easily be taken
into account). The free parameter ${\epsilon}$ controls the
fraction of the kinetic energy of supernovae that is available to
power the winds, and $v_{\rm esc}^{2}(z)$ is the mean square of
the escape velocity of structures at redshift $z$.

In order to compute the stellar lifetime $\tau_s(m)$ we assume it
to be equal to the time that a star of mass $m$ spends on the main
sequence. So the age of a star of mass $m$ is given by
\begin{equation} \label{eq:taus}
\tau_s(m) = (m/M_{\odot})^{-2.5}t_{\odot}
\end{equation}
where $t_{\odot}$ is the total time that a star of mass
$M=M_{\odot}$ will spend on the main sequence and we adopt a value
$t_{\odot}=9$~Gyr. To compute $m_d(t)$ in Eq.~\eqref{eq:m0} we
solve Eq.~\eqref{eq:taus} for $m$, thereby obtaining the mass of
stars $m_d(t)$ that die at age $t$.

The escape velocity is obtained by assuming virialized halos and
averaging over the distribution function, thus
obtaining\footnote{This is the escape velocity from $R$ to
infinity, not the escape velocity from the sites of star formation
that are deeper in the potential well. This approximation does not
affect the results for the massive spheroids since even the
shallower potential at $R$ is deep enough to prevent winds from
being effective, see the discussion in section~\ref{sec:results}.
For less massive systems we expect this approximation to result in
slightly smaller values for the parameter $\epsilon$ than one
would otherwise obtain.}:
\begin{equation}
v_{\rm esc}^{2}(z)~=~\frac{{\int}_{M_{\rm min}}^{\infty}dMMf_{\rm
ps}(M,z)(2GM/R(M))}{{\int}_{\rm M_{min}}^{\infty}dMMf_{\rm
ps}(M,z)}
\end{equation}
where $R(M)$ is the radius of a dark matter halo of mass $M$ given
by the following expression:
\begin{equation}
R(M)~=~\left(\frac{3M}{178{\rho}_{\rm c}({\Omega_{\rm
m}}(1+z)^{3}+{\Omega}_{\rm \Lambda})4{\pi}}\right)^{1/3},
\end{equation}
where ${\rho}_{\rm c}$ is the critical density of the universe
today.The factor $178$ is the overdensity (relative to the
critical density) at virialization for an Einstein--de Sitter
model~\cite{coles95}.

\subsection{Star formation and supernova rate}

Following \pcite{daigne05}, we adopt an exponentially decreasing
SFR:
\begin{equation}
\Psi(t) = \nu \Mstruc(t)\exp(-(t-t_{\rm init})/\tau),
\label{eq:star_form}
\end{equation}
where $t_{\rm init}$ is the time corresponding to the redshift
$z_{\rm init}$ when star formation starts (as defined above),
$\tau$ is a characteristic time scale that we take as a free
parameter and $\nu$ is a normalization parameter (with dimensions
of inverse time).

The supernova rate (SNR) is strongly linked to the star formation
rate because of the short lifetime of massive progenitors with
$M>8M_{\odot}$. We can therefore assume that core collapse
supernovae (SNe) are strongly correlated with instantaneous SFR,
and the supernovae type II rate ${\Psi}_{\rm sn}(t)$ is given by
\begin{equation}
\label{eq:SNR} {\Psi}_{\rm sn}(t) = \int_{8M_\odot}^{m_{\rm
u}}{\Phi}(m){\Psi}(t-{\tau_s(m)})dm.
\end{equation}

\subsection{Chemical evolution model}
\label{sec:Chemical Evolution}

Chemical evolution is included in the model using the
instantaneous recycling approximation. i.e. we assume that all
processes involving stellar evolution, nucleosynthesis and
recycling take place instantaneously on the timescale of galactic
evolution. The equation of galactic chemical evolution is
\cite{pagel97}
\begin{equation}
\Sigma_{\rm g}\frac{dZ}{dt} = q{\Psi}(t)+(Z_{F}-Z(t))a_{\rm
b}(t)-({\eta}-1)Z(t)o(t), \label{eq:chem_enrich}
\end{equation}
where $\Sigma_{\rm g}$ is the density of the gas~(in units of
$M_{\odot}/\text{Mpc}^{3}$), $\eta$ is a multiple of the
nucleosynthetic yield that parameterises the metallicity $Z$ of
the SN ejecta~\cite{dalcanton06} (also called ``the load factor'',
adopted here as a free parameter) and $q$ is the yield. We fix the
value of the yield to $q=0.02$ and assume that the mass accreted
to the disk has zero metallicity, i.e. we fix $Z_{F} = 0$).
Furthermore, we normalise the metallicities to the solar value for
which we adopt $Z_{\odot}=0.02$. The chemical evolution of the
ISM, described by Eq.~\eqref{eq:chem_enrich} contains three terms.
The first one represents the chemical enrichment due to the
evolution of stars. The second term represents the dilution of
metallicity (if $Z_{F} < Z(t)$) or the chemical enrichment (if
$Z_{F} > Z(t)$) of the ISM due to accreted material. The last term
describes the dilution of metallicity (if ${\eta} > 1$) or the
chemical enrichment (if ${\eta} < 1$) of the ISM due to galactic
winds powered by stellar explosions.

In recent theoretical work, what has been dubbed ``the missing
metals problem'' has received considerable attention (see
\pcite{prochaska03} for an extensive discussion of this problem),
namely the fact that the mean metallicity is $\sim 10$ times lower
than the value expected from the inferred star formation history.
This problem may indicate a serious flaw in our understanding of
the interplay between star formation and metal enrichment.
Therefore, we have introduced in our chemical evolution model an
extra parameter $f_{\rm dil}$ accounting phenomenologically for
these effects. This allows the metallicity predictions of
Eq.~\eqref{eq:chem_enrich} to be adjusted to match observational
data. Thus we rescale the metallicity values given by solving
Eq.~\eqref{eq:chem_enrich} by a factor $f_{\rm dil}$, i.e.
\begin{equation}
\label{eq:fdil} {\tilde{Z}(t)} = Z(t)/f_{\rm dil}.
\end{equation}

\subsection{Summary of model parameters}
\begin{table*}
\begin{minipage}{177mm}
\centering
\begin{tabular}{l l l l l }
 \hline \\
 Quantity & Symbol & Defined & Prior range or value\\\hline
 Minimum mass of collapsed haloes ($M_{\rm min}$ in $M_\odot$) & $\logM$
 & Sec.~\ref{sec:acc} &
 $5\leq \logM \leq 13$
 \\
 SN type II energy efficiency factor & $\epsilon$ & Eq.~\eqref{eq:eps} & $0.01\leq\epsilon\leq 0.45$\\
 IMF power--law index&   $x$ &  Eq.~\eqref{eq:x} & $3\leq x \leq 2$\\
 SFR normalization parameter (Gyr$^{-1}$)& $\nu$ & Eq.~\eqref{eq:star_form} & $0.01\leq \nu \leq 5$\\
 SFR timescale (Gyr) & $\tau$& Eq.~\eqref{eq:star_form} & $1\leq\tau\leq 5$\\
 Winds load factor& $\eta$ & Eq.~\eqref{eq:chem_enrich} & $0\leq\eta\leq 30$ \\
 Metals dilution factor & $f_{\rm dil}$ & Eq.~\eqref{eq:fdil} & $1\leq\fdil\leq 30$\\\hline
 Baryon density parameter & $\Omega_b$ & & $0.044$ \\
 Matter density parameter & $\Omega_m$  & & $0.27$  \\
 Cosmological constant density parameter & $\Omega_\Lambda$ & & $0.70$  \\
 Hubble constant (km/sec/Mpc) & $H_0$  & & $71$  \\
 Rms fluctuations amplitude  & $\sigma_8$  & & $0.9$  \\
 Dark matter to baryons bias parameter & $b$  & & $1.0$  \\
 Minimum fraction of baryons when star formation begins & $f_\text{bar}(z_\text{init})$  & & $0.01$  \\
 Kinetic energy from stellar explosions & $E_\text{kin}$ & & $10^{51}$ ergs\\
 Yield & $q$ & & $0.02$\\
 Metallicity of accreted material & $Z_F$ & & $0$ \\\hline
   \end{tabular}
\caption{Upper part: free model parameters and priors used in the
analysis. Top--hat (flat) priors have been adopted on the
parameter ranges indicated. Lower part: model parameters that have
been fixed.} \label{tab:params}
\end{minipage}
\end{table*}

To summarize, our model is characterized by a set of 7 free
parameters, that we denote by $\params$:
 \begin{equation}
 \label{eq:params}
 \params = \left( \logM, \epsilon, x, \nu, \tau, \eta, \fdil \right)
 \end{equation}
The free parameters of the model (and the ones that we have chosen
to fix) are summarized in Table~\ref{tab:params}, where we also
give the prior ranges for our statistical analysis, i.e. the
ranges within which their values are allowed to vary (see
\ref{sec:mcmc} for more details).

\section{Influence of model parameters on observable quantities}
\label{sec:Sensitivity}

In this section, we discuss the impact of each of the 7 free
parameters in our model (given in Eq.~\eqref{eq:params} and in the
upper part of Table~\ref{tab:params}) on the physical observables
introduced above, namely the SFR, SN type II rate, metallicity and
baryonic fraction in structures. We also present a physical
interpretation of the observed behaviour of these quantities. As a
fiducial model we fix the parameter values to the following
values: $\logM=8$, $\epsilon=0.1$, $x=1.7$, $\tau=3$, $\nu=1.4$,
$\eta=10$ and $\fdil=2$. We then proceed to vary one of the
parameters at a time to get a feeling for the physical impact of
each of them.

\begin{figure}
\centering
 \includegraphics[width=1.0\linewidth]{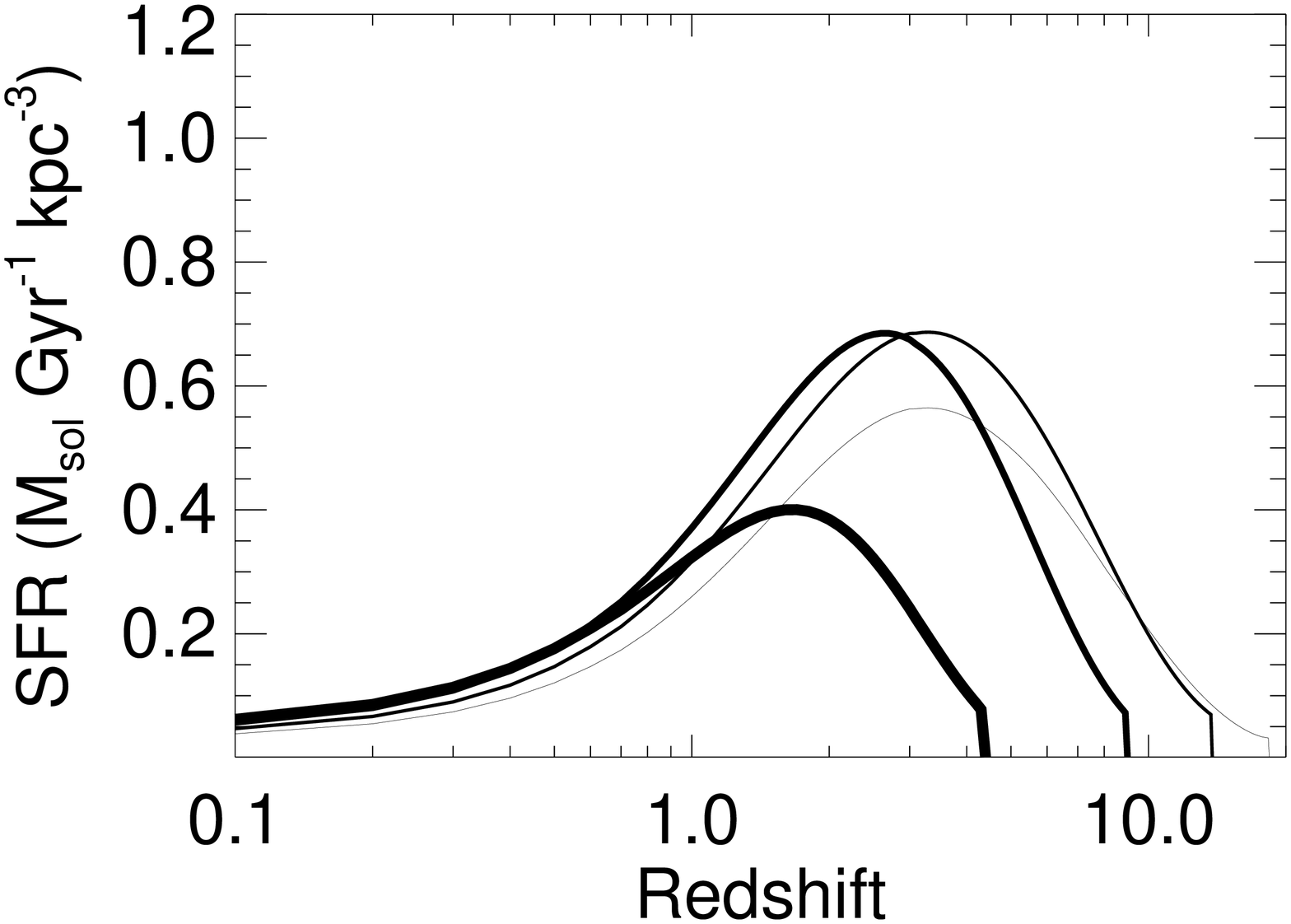}\\
 \includegraphics[width=1.0\linewidth]{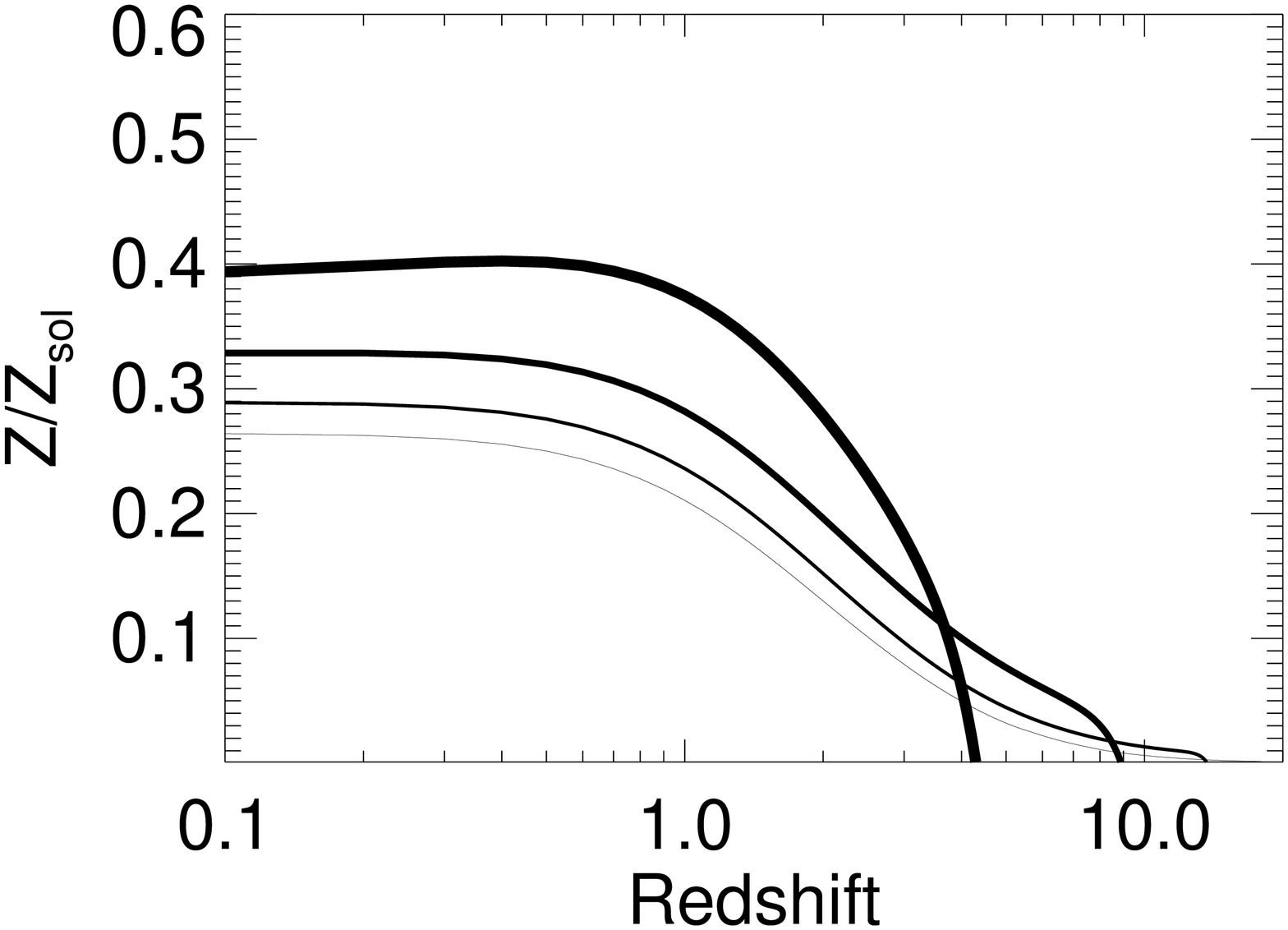}\\
 \includegraphics[width=1.0\linewidth]{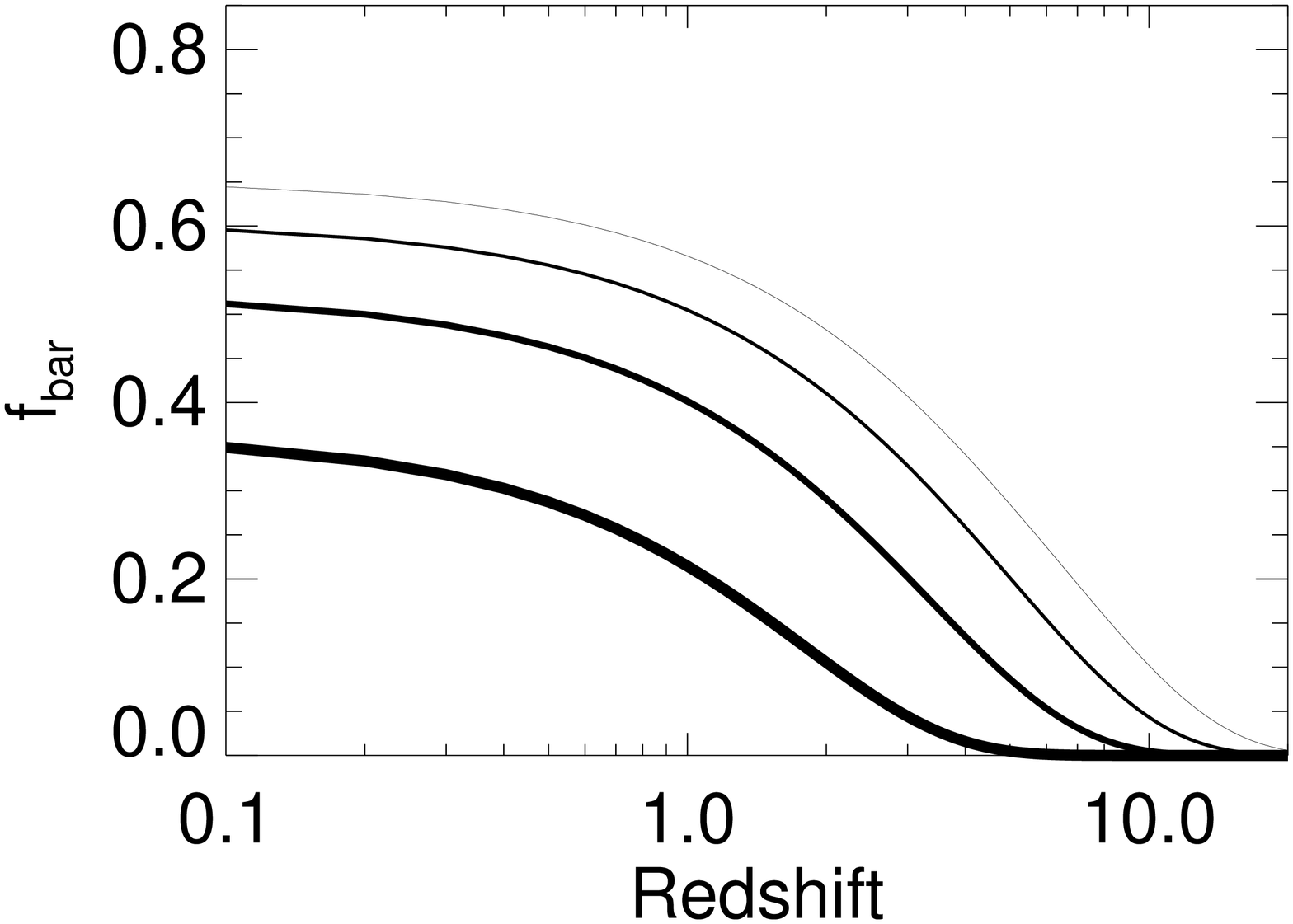}
\caption{Dependence of the SFR, metallicity and baryonic fraction
in structures (panels from top to bottom) on the minimum mass of
collapsed dark matter halos, $\logM$. The curves are for
$\logM=6,8,10,12$, from thin to thick.\label{fig:Sen_M}}
\end{figure}

Figure~\ref{fig:Sen_M} shows the model dependence on the minimum
mass of collapsed dark matter halos, $\logM$. Smaller values of
this parameter describe a scenario where star formation is
hierarchical and follows the growth of structures, while higher
values of $\logM$ correspond to star formation occurring in
massive spheroids. Correspondingly, for small $\logM$ star
formation begins earlier, as apparent from the top panel of
Figure~\ref{fig:Sen_M}. At small redshift, a smaller $\logM$ leads
to reduced SFR, since the relatively strong winds ($\epsilon=0.1$
in this example) drive the gas out of the system for shallower
potentials. For large $\logM$, the build--up of metals is delayed
in time but the metallicity can reach larger values, since
supernova--powered winds are less important in massive systems
(middle panel). As it is clear from Eq.~\eqref{eq:Bar_frc}, the
percentage of baryons allocated to dark matter halos $f_{\rm bar}$
increases for decreasing $\logM$ (bottom panel). Due to the short
lifetime of massive progenitors, the SN rate is essentially
identical to the SFR, and we therefore do not display it.

\begin{figure}
\centering
 \includegraphics[width=1.0\linewidth]{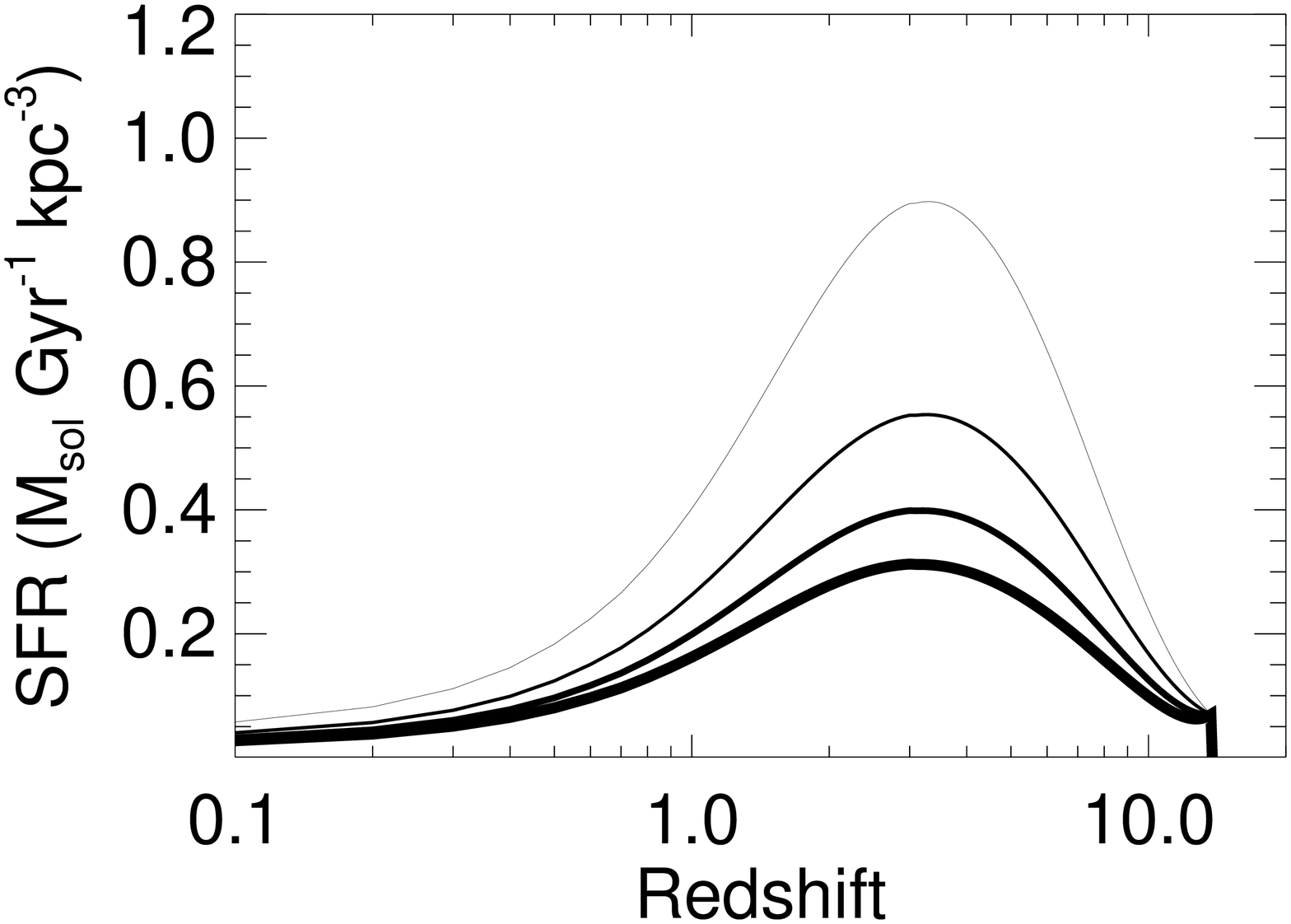}\\
 \includegraphics[width=1.0\linewidth]{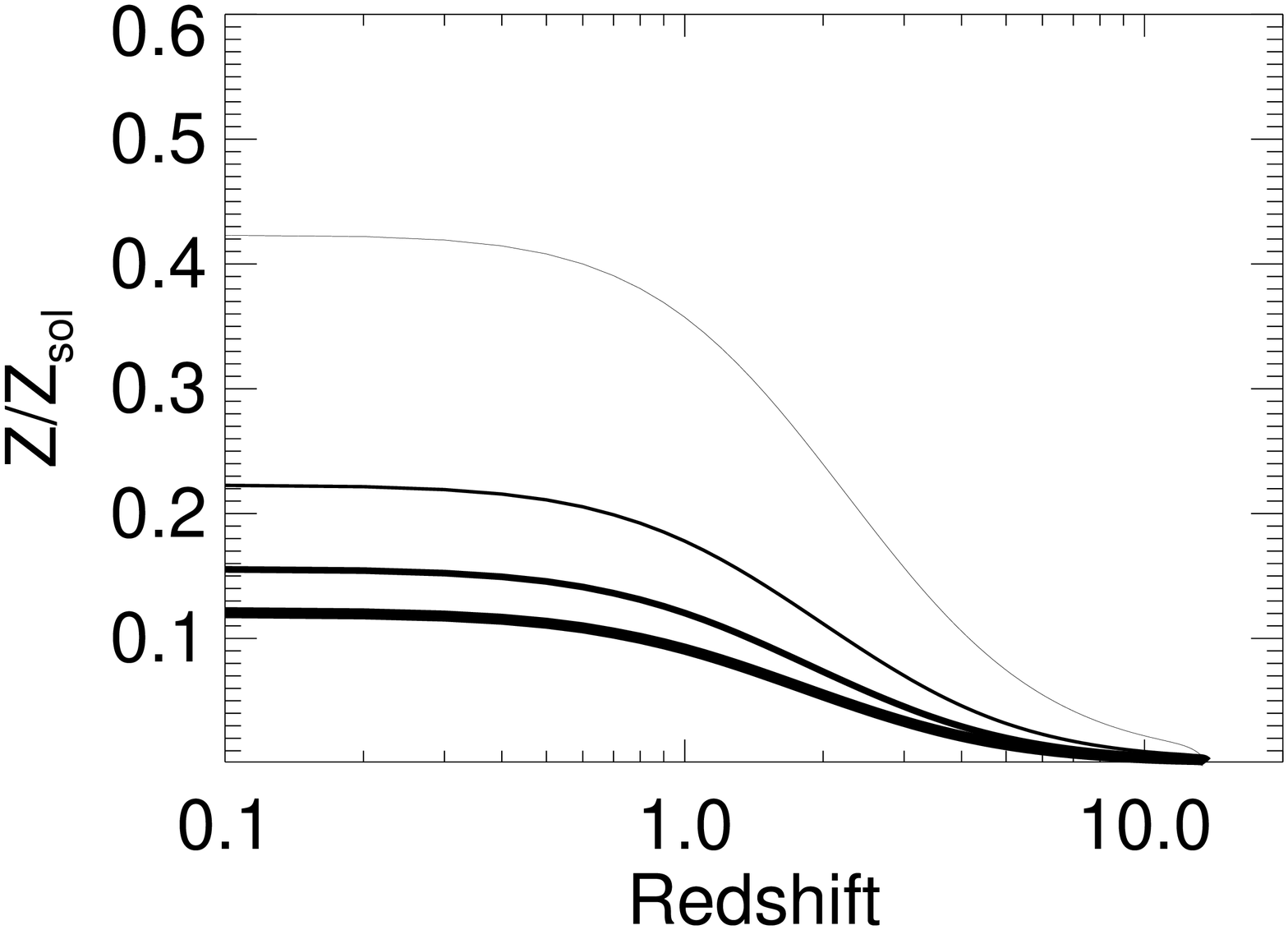}\\
\caption{Dependence of the SFR (top panel) and of the metallicity
(bottom panel) on the model parameter $\epsilon$, describing the
strength of galactic winds. The curves are for $\epsilon=0.05,
0.15, 0.25, 0.35$, from thin to thick.} \label{fig:Sen_E}
\end{figure}

In Figure~\ref{fig:Sen_E} we show the model sensitivity to the
parameter $\epsilon$, defining the percentage of supernovae energy
that goes to the ISM. This parameter essentially describes the
strength of galactic winds. The physical interpretation of high
values of $\epsilon$ is that strong winds driven by feedback
energy are maintained in dark matter halos. For increasing value
of $\epsilon$, galactic winds become stronger and the star
formation rate is reduced since less gas is available to make
stars (top panel). This effect is more important for the shallower
gravitational potential of the low mass halos, i.e. for smaller
$\logM$ (in this example, $\logM = 8$). As already remarked above,
higher values of $\epsilon$ corresponds to winds sweeping out
metals from the ISM and thus to lower metallicity (bottom panel).
Again, this effect is most important for the low mass halos where
their gravitational potential is relatively shallow.

\begin{figure}
\centering
 \includegraphics[width=1.0\linewidth]{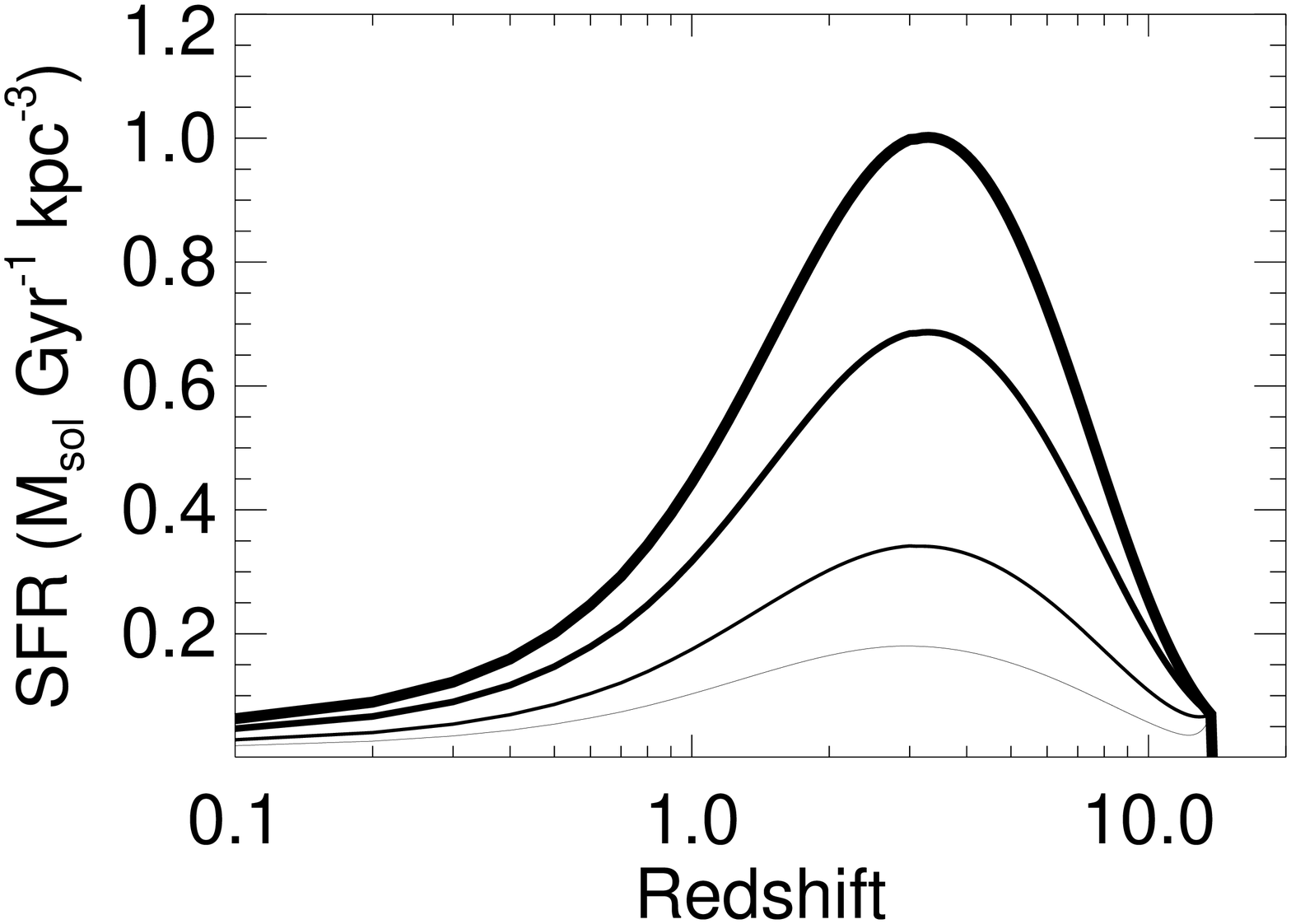}\\
 \includegraphics[width=1.0\linewidth]{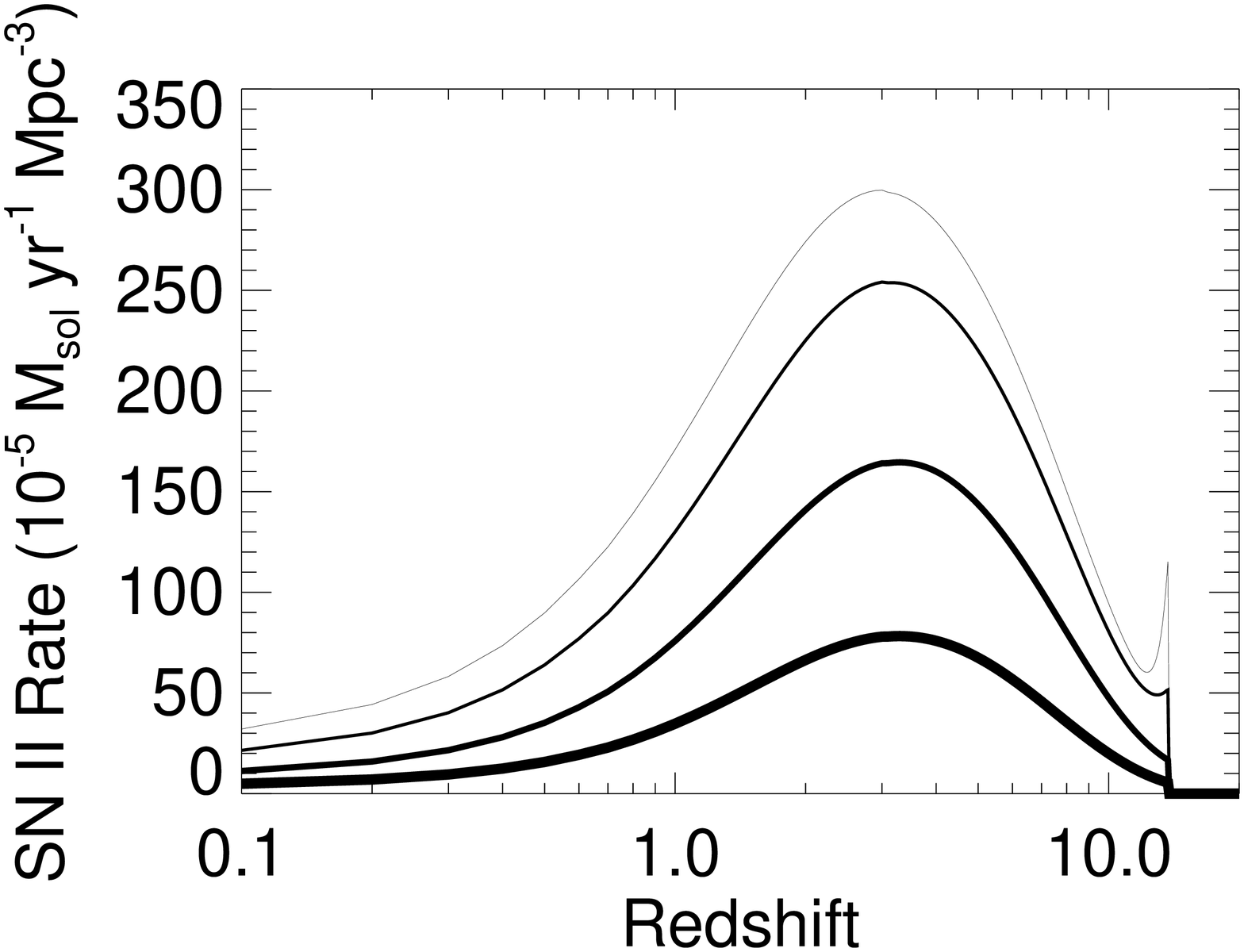}\\
 \includegraphics[width=1.0\linewidth]{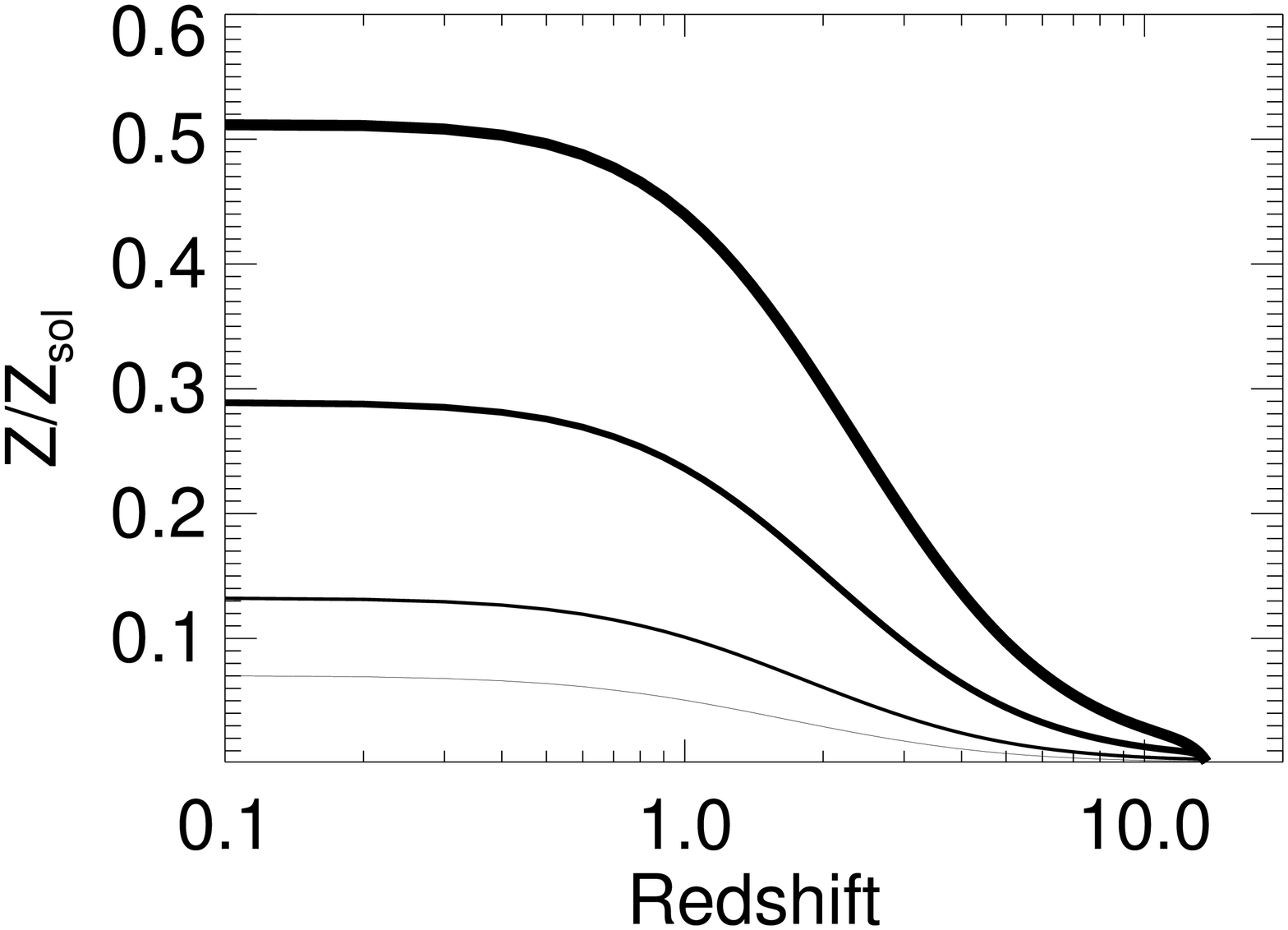}\\
\caption{Dependence of the SFR (top panel), SN rate (middle panel)
and of the metallicity (bottom panel) on the model parameter $x$,
controlling the slope of the IMF. Curves are for $x=1.0, 1.4, 1.7,
2.0$, from thin to thick. } \label{fig:Sen_X}
\end{figure}

The sensitivity to the parameter $x$, giving the slope of the
initial mass function, is shown in Figure~\ref{fig:Sen_X}. We can
see a strong influence of $x$ on the supernovae type II rates
(middle panel), a consequence of Eq.~\eqref{eq:SNR}. Decreasing
the value of $x$ (i.e., making the IMF shallower) corresponds to a
larger number of more massive stars, and hence the supernovae type
II rate increases. Taking into account that in our model each
supernova gives a constant percentage of its energy to the ISM,
small values of $x$ result in stronger galactic winds and so in
smaller star formation rates (top panel) and a less enriched ISM,
hence smaller metallicity (bottom panel). For the extreme case
that $x=1$~(very flat IMF) the supernovae type II rate is very
large at high redshift causing very strong winds that reduce the
SFR quickly. This causes the spike in
Figure~\ref{fig:Sen_X}~(middle panel).

\begin{figure}
\centering
 \includegraphics[width=1.0\linewidth]{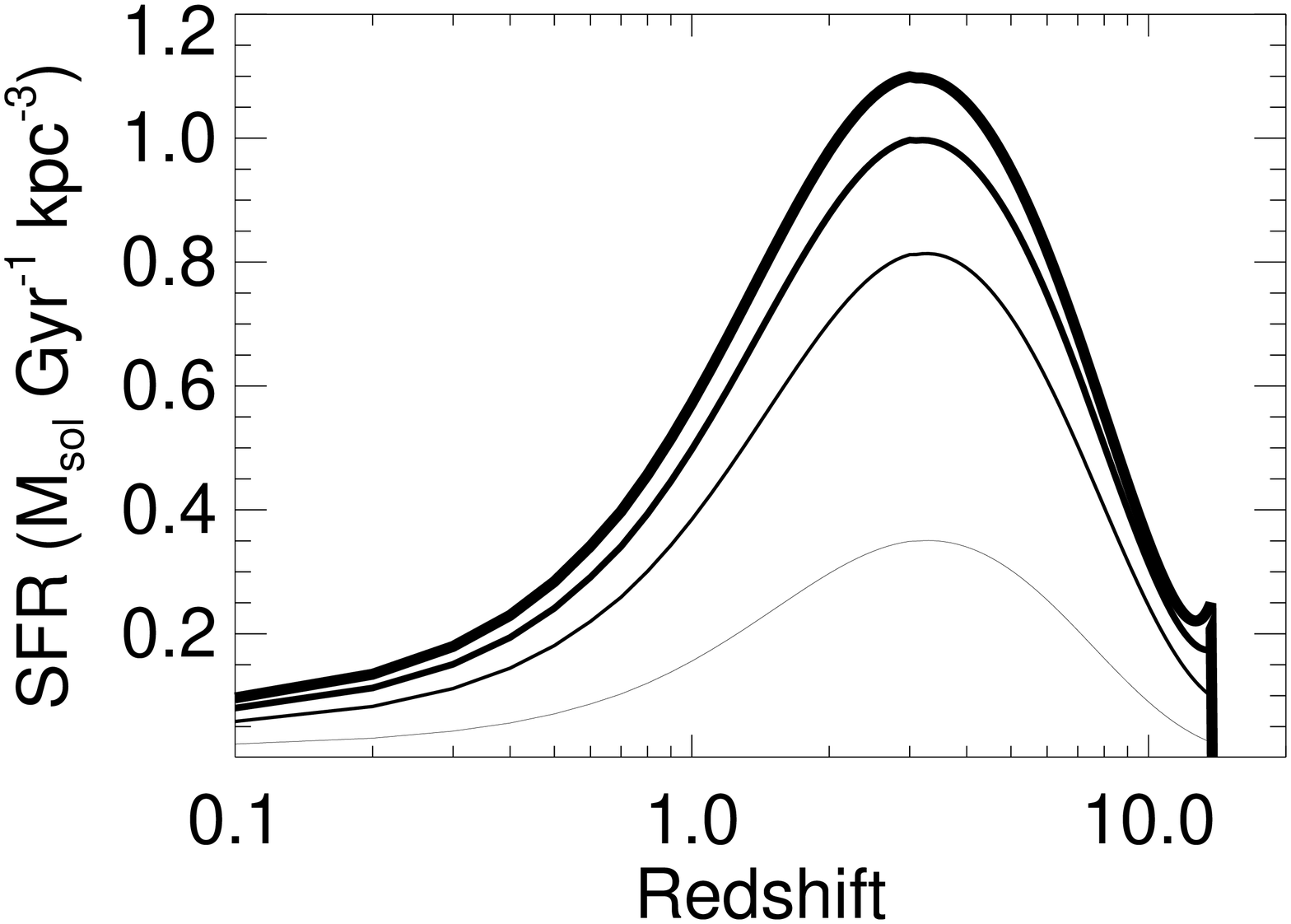}\\
 \includegraphics[width=1.0\linewidth]{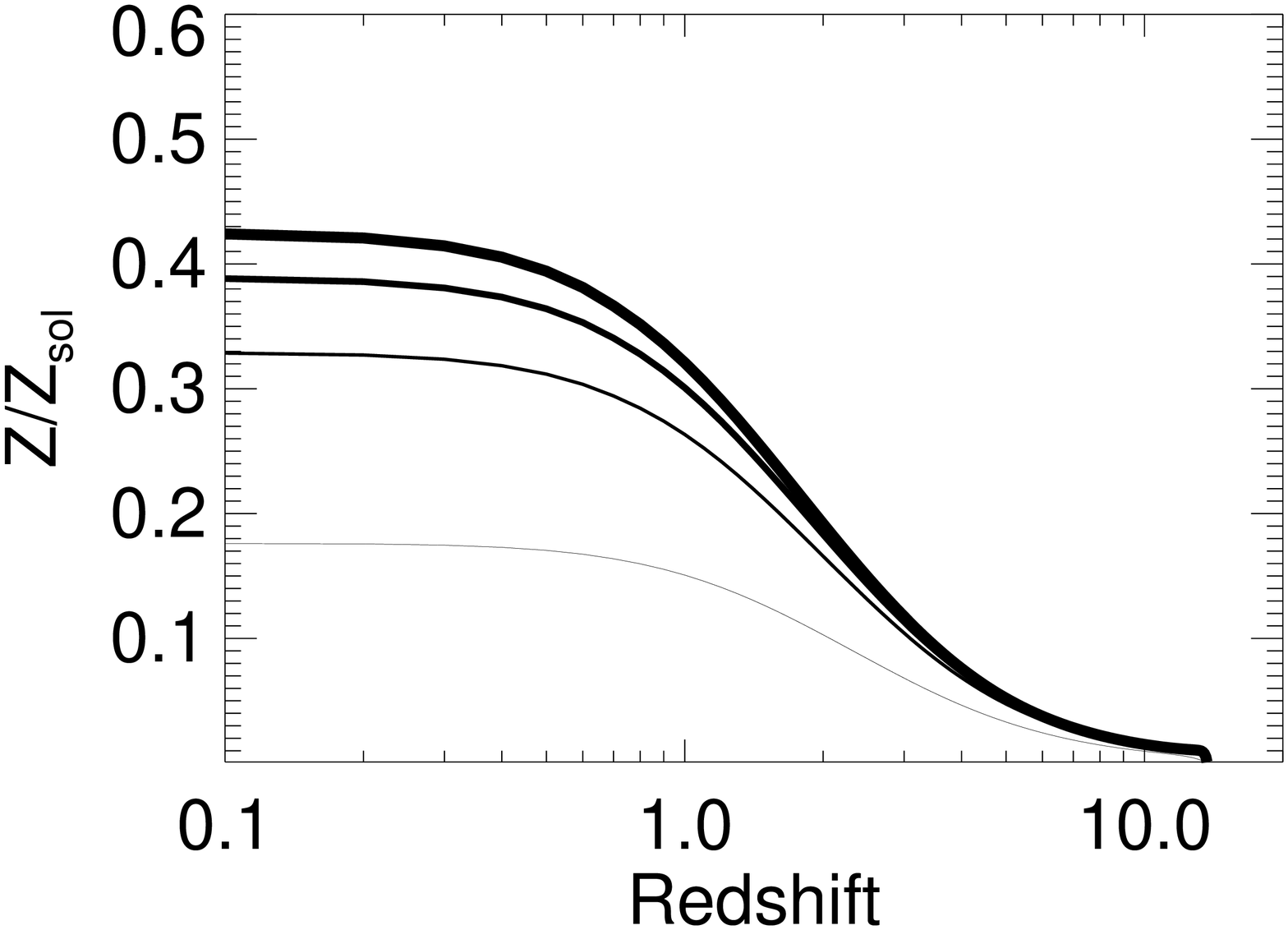}\\
\caption{Dependence of the SFR (top panel) and the metallicity
(bottom panel) on the parameter $\nu$, controlling the efficiency
of star formation. The curves are for $\nu=0.5, 2, 3.5, 5$, from
thin to thick.} \label{fig:Sen_V}
\end{figure}

Figure~\ref{fig:Sen_V} shows the model sensitivity to the
parameter $\nu$, connected with star formation through the
proportionality factor that defines the efficiency of star
formation, see Eq.~\eqref{eq:star_form}. Increasing the value of
$\nu$ the star formation becomes more efficient and the
interstellar medium becomes highly enriched in metals by evolving
stars. On the contrary, smaller values of $\nu$ lead to a less
efficient star formation.

\begin{figure}
\centering
 \includegraphics[width=1.0\linewidth]{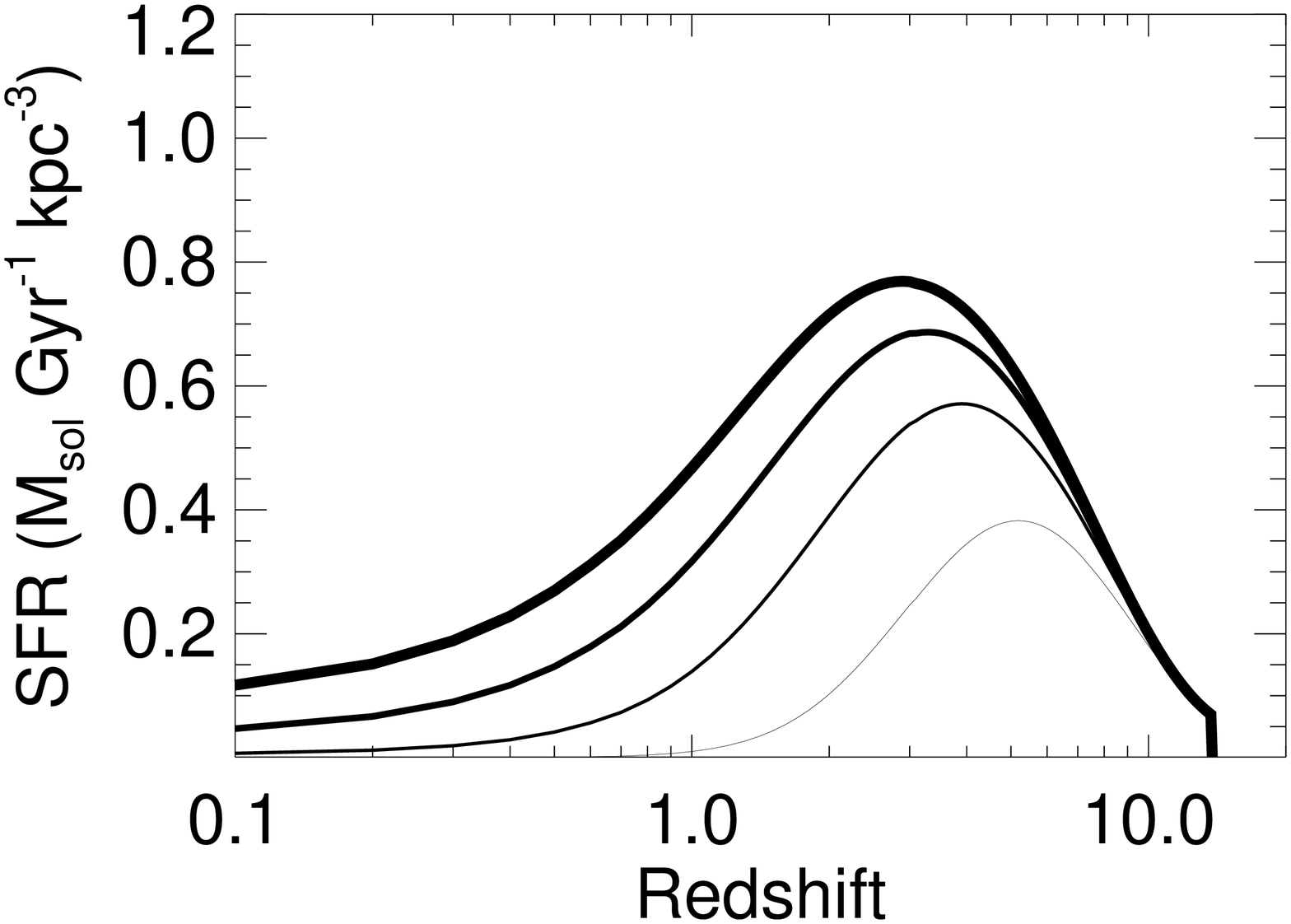}\\
 \includegraphics[width=1.0\linewidth]{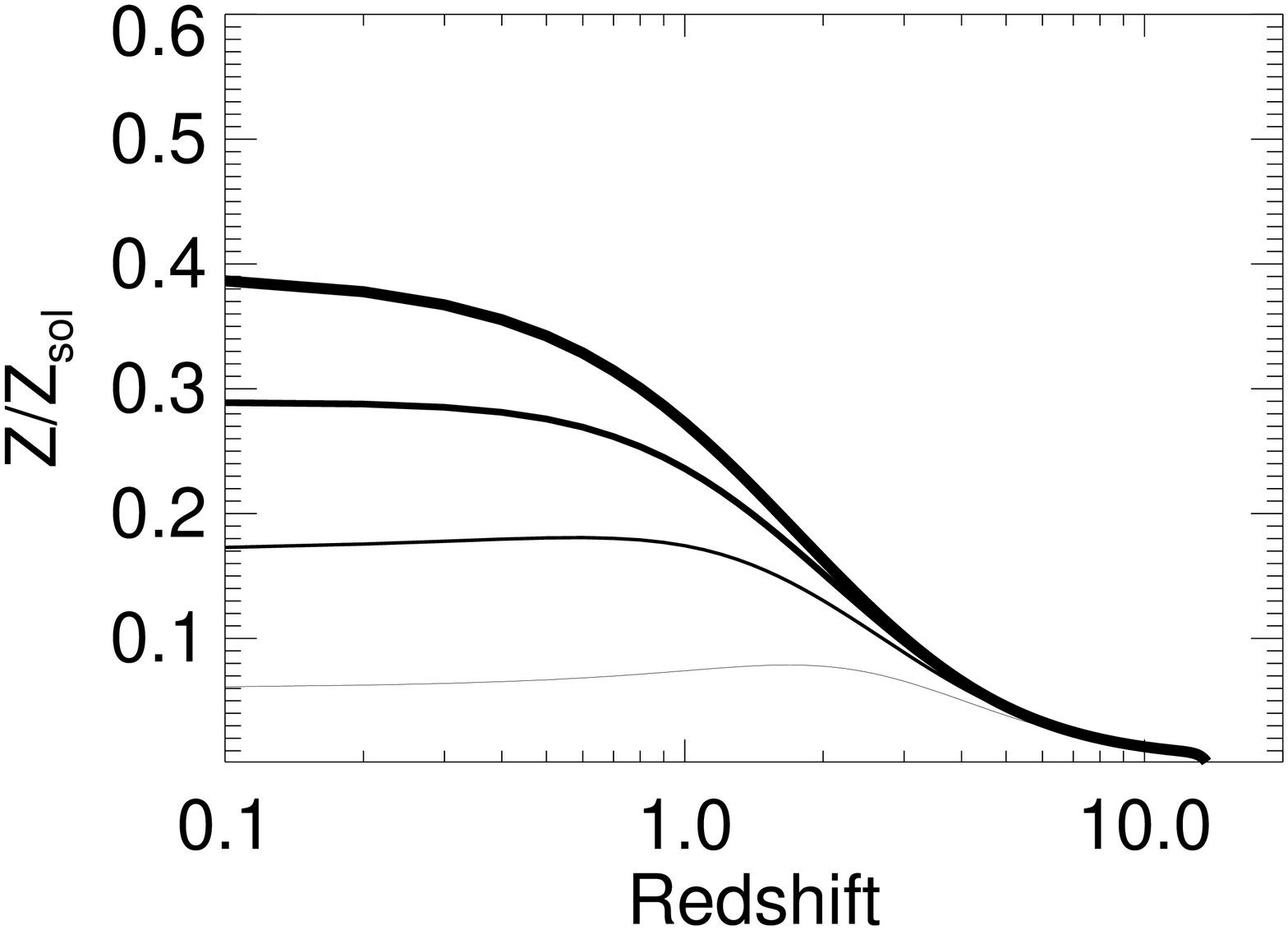}\\
\caption{Sensitivity of the model to the parameter $\tau$, giving
the characteristic timescale of star formation. The curves are for
$\tau=1, 2, 3, 4$, from thin to thick.} \label{fig:Sen_T}
\end{figure}

The influence of the parameter $\tau$, defining the characteristic
timescale of star formation, is displayed in
Figure~\ref{fig:Sen_T}. Decreasing the value of $\tau$ leads to
the star formation activity ending sooner and to an ISM which is
therefore poorer in metals. Larger values of $\tau$ result in an
enriched ISM since galaxies are active, in terms of star
formation, for a longer period.

\begin{figure}
\centering
 \includegraphics[width=1.0\linewidth]{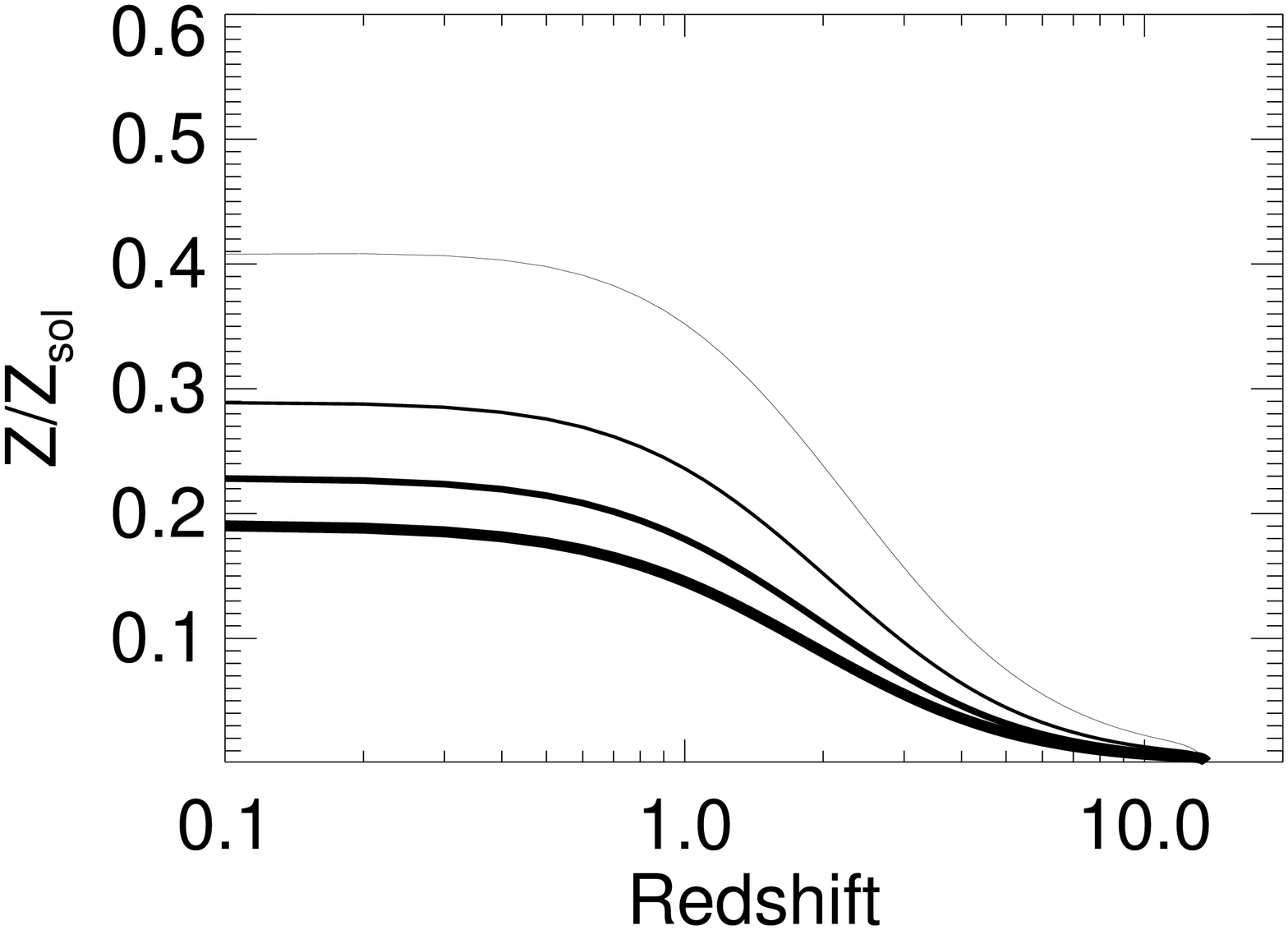}
\caption{Sensitivity of the metallicity to the parameter $\eta$.
Curves are for $\eta=5,10,15,20$, from thin to thick.}
\label{fig:Sen_eta}
\end{figure}

The influence of the parameter $\eta$, controlling the metallicity
of the ejecta, is displayed in Figure~\ref{fig:Sen_eta}. A larger
value of $\eta$ leads to a decrease in metallicity of the system,
since the metallicity of the winds is increased by a factor of
$\eta$ wrt the mean metallicity, see Eq.~\eqref{eq:chem_enrich}.
Finally, we do not display the impact of the dilution factor
$\fdil$, since its value merely rescales the metallicity by a
multiplicative factor, see Eq.~\eqref{eq:fdil}.

We now turn to discuss the data employed and the details of our
statistical treatment and fitting procedure.

\section{Data and statistical analysis}
\label{sec:data}

Combining different types of observations to maximise their
constraining power on multi--dimensional parameter spaces has
become a common approach in cosmology. Following an approach
similar in spirit, in this work we perform a simultaneous analysis
of star formation history, SN rates, metallicity and baryonic
fraction data in order to find tight constraints on the parameters
of our model, Eq.~\eqref{eq:params}. One of the aim of this paper
is to provide the first complete statistical analysis of existing
metallicity, SFR, SNII rate and local collapse baryon fraction
data in a realistic model. We first describe the data employed in
section \ref{sec:stat_treat}, then we outline the Bayesian fitting
procedure that vastly improves on usual fixed--grid scans in
section \ref{sec:mcmc}.

\subsection{Observational constraints}
\label{sec:stat_treat}

The usual compilations of measurements for the SFR and metallicity
observations (such as the ones used e.g. in \pcite{daigne05} are
unsuitable for a robust statistical analysis, because of the large
systematic differences among measurements at about the same
redshift performed over a range of different systems. In fact,
when using such a ``raw'' data compilation the statistical fit is
usually dominated by only a few data points with very small
errorbars, while the large majority of observations carry almost
no statistical weight. This is clearly less then satisfactory. To
cure this effect, it becomes important to bin the observations in
such a way as to account for possible systematic uncertainties
among different measurements at the same redshift. This problem is
addressed here for the first time by employing a Bayesian
procedure that accounts for possible systematic differences
between measurements, based on the treatment given
in~\cite{press96}. The details of the method are given in
Appendix~\ref{app}.

We apply the binning procedure described in the Appendix to the
data points for the metallicity in the interstellar medium (ISM)
given by~\pcite{prochaska03}. By using Eq.~\eqref{eq:prob_bin} we
place the 125 measurements in 8 bins, ranging in redshift from $z
= 0.85$ to $z = 4.45$. The bin distribution and spacing has been
chosen to obtain a reasonable large number of points in each bin,
while simultaneously having a sufficiently small redshift spacing
between bins. The measurements of [M/H] number density relative to
solar metallicity obtained after the statistical rebinning, are
summarized in Table~\ref{Table.Cosmic_Metal}.

\begin{table}
  \begin{center}
  \begin{tabular}{|l l l|}
    \hline
    Redshift& Metallicity&  Number \\
    $z_b$& [M/H]/[M/H]$_{\odot}$&  of points\\
    \hline
     0.85 &$-0.83 \pm {0.11}$&6\\
     1.45 &$-1.06 \pm {0.09}$&4\\
     1.95 &$-0.93 \pm {0.12}$&17\\
     2.45 &$-1.36 \pm {0.14}$&29\\
     2.95 &$-1.56 \pm {0.19}$&18\\
     3.45 &$-1.78 \pm {0.08}$&28\\
     3.95 &$-1.80 \pm {0.06}$&16\\
     4.45 &$-1.76 \pm {0.11}$&7\\
    \hline
  \end{tabular}
\end{center}
\caption{Binned measurements of [M/H] number density relative to
solar values, after the statistical treatment of the data (see
Appendix~\ref{app_noz} for details).} \label{Table.Cosmic_Metal}
\end{table}

For the case of cosmic SFR data, our statistical rebinning is
modified in order to take into account the redshift uncertainty in
the raw data. Details are given in Appendix~\ref{app_z}. We take
the compilation of ``raw'' data out to $z \sim 5$ from
\pcite{hopkins04}, excluding only one measurement corresponding to
the cosmic star formation at $z =0.005{\pm}0.005$, reported by
\pcite{condon89}. Instead, we replace this point by more recent
measurement at the same redshift as reported by the same
author~\cite{condon02}. Both these measurements use as cosmic star
formation estimator counts at $1.4$~GHz. From the raw data we
derive binned values in 12 redshift bins, with centers ranging
fron $z=0.035$ to $z=5.12$ by using~Eq.~\eqref{eq:prob_bin_z}. The
resulting bins with their errors are summarized in
Table~\ref{Table.Cosmic_Sfr}.

\newcommand{\hi}{\rule{0pt}{14pt}}
\begin{table}
  \begin{center}
  \begin{tabular}{|l l |}
    \hline
    Redshift& SFR density   \\
    $z_b$           & $\log(\dot{\rho}_\star) [M_{\odot}\text{yr}^{-1}\text{Mpc}^{-3}]$ \\
    \hline
     0.012    &$-1.78 \pm 0.18$\\
     0.135    &$-1.45 \pm 0.07$ \\
     0.275    &$-1.45 \pm 0.06$ \\
     0.405    &$-1.37 \pm 0.23$ \\
     0.580    &$-1.08 \pm 0.08$\\
     0.755    &$-1.03 \pm 0.06$\\
     0.905    &$-0.98 \pm 0.07$\\
     1.150    &$-0.92 \pm 0.09$ \\
     1.650    &$-0.63 \pm 0.26$\\
     2.520    &$-0.63 \pm 0.21$\\
     3.770    &$-0.79 \pm 0.10$\\
     5.120    &$-0.88 \pm 0.29$\\
    \hline
  \end{tabular}
\end{center}
\caption{SFR density data after our statistical binning of the the
``raw'' SFR data compilation (see Appendix~\ref{app_z} for
details). No dust correction has been applied to this values. }
\label{Table.Cosmic_Sfr}
\end{table}

Furthermore, the SFR predictions of our model are corrected to
account for dust absorption. There are large uncertainties
associated with dust absorption correction, this is why at low
redshift (i.e., for bins with $z_b\leq 3$) we employ both a
``normal dust correction'' of 1.0 mag and a ``large dust
correction'' of 1.8 mag. These two choices are made in view of the
fact that they seem to bracket the expected values valid over a
broad range of systems~\cite{schiminovich05}. For bins at a higher
redshift ($z_b>3$) we adopt a fixed dust correction of 0.4 mag,
following~\pcite{schiminovich05}. We shall see in the next section
that the dust absorption correction scheme one adopts has a
crucial impact on the resulting physical scenario.

The present--day fraction of baryons in structures, as estimated
by ~\pcite{fukugita04} is taken to be
\begin{equation}
 \label{eq:bardata}
f_{\rm bar}(z = 0) = 0.61 {\pm} 0.11.
\end{equation}

The data for the core collapse supernovae are taken from the Great
Observatories Origins Deep Survey~(GOODS,~\pcite{dahlen04}). The
GOODS core collapse supernovae rates have been placed in two bins
at $z=0.3{\pm}0.2$ and $\rm z=0.7{\pm}0.2$. For the local rate (at
$z=0$) we adopt the value from \pcite{cappellaro99}. We convert
the local rate from supernovae units as described in
\pcite{dahlen04}. The 3 above mentioned data points are summarized
in Table~\ref{Table.Cosmic_Snr}.

\begin{table}
  \begin{center}
  \begin{tabular}{|l l|}
    \hline
    Redshift& SN type II rate \\
    $z$           &  $[\text{Sne yr}^{-1}\text{Mpc}^{-3}]$ \\
    \hline
     0.0 & $6.16 \pm 2.92$ \hi\\
     0.3 & $26.20_{-9.18}^{+7.83}$\hi\\
     0.7 & $41.32_{-10.75}^{+11.06}$\hi\\
    \hline
  \end{tabular}
\end{center}
\caption{Measurements of supernovae type II rate as a function of
redshift.} \label{Table.Cosmic_Snr}
\end{table}

\subsection{Bayesian Markov Chain Monte Carlo analysis}
 \label{sec:mcmc}

After the statistical rebinning of the data described above, the
likelihood function $P(d|\params)$ is the sum of four independent
terms, describing the observations of the SFR, the metallicity,
the SN rate and the baryonic fraction, respectively:
 \begin{equation}
 P(d|\params) =  {\cal L}_{\text{SFR}} + {\cal L}_{\text{met}} + {\cal
 L}_{\text{SN}}+ {\cal L}_{b}.
 \end{equation}
We model each of the above four terms as a product of the data
points for each observable, taken to be independent and with
Gaussian noise
\begin{equation} \label{eq:likelihood}
\chi^2_\text{obs} =  - 2 \ln {\cal L}_{\text{obs}} =
\sum_{i=1}^{N_\text{obs}}\frac{(y_i - d_i)^2}{\sigma_i^2}
\end{equation}
where ``obs'' stand for SFR, metallicity, SN or baryon fraction in
structures, and the means $d_i$ and standard deviations $\sigma_i$
of the data points are given in Tables
\ref{Table.Cosmic_Metal}--\ref{Table.Cosmic_Snr} and
Eq.~\eqref{eq:bardata}. The normalization constant does not
matter, as we are only interested in the relative posterior
probability density, as we now discuss.

From the likelihood function of Eq.~\eqref{eq:likelihood} we
obtain the posterior probability for the parameters of interest,
$P(\params|d)$, via Bayes' theorem,
 \begin{equation}
 \label{eq:bayes}
 P(\params|d) = \frac{P(d | \params)P(\params)}{P(d)},
 \end{equation}
where $P(\params)$ is the prior probability distribution
(``prior'' for short) and $P(d)$ is a normalization constant that
does not depend on the parameters and can therefore be neglected
in the following (see \pcite{Trotta:2005ar,Trotta:2007hy} for more
details on Bayesian parameter inference and model comparison). We
adopt flat (i.e., top--hat) priors on our set of parameters
$\params$ given in Eq.~\eqref{eq:params} in the ranges given in
Table~\ref{tab:params}, which means that the posterior probability
distribution function (pdf) is simply proportional to the
likelihood.

In order to explore efficiently our our 7--dimensional parameter
space, we employ a Markov Chain Monte Carlo (MCMC) procedure, with
some of the routines adapted from the publicly available {\sc
cosmomc} package\footnote{Available from
\texttt{cosmologist.info}}. The great advantages of MCMC methods
are that the computational time scales approximately linearly with
the number of dimensions of the parameter space, and that the
marginalized posterior distribution for the parameters of interest
and their correlations can be simply recovered by plotting
histograms of the sample list. We follow the procedure outlined in
\pcite{de Austri:2006pe}, to which we refer for further details.
Here we only briefly sketch the main points.

The aim of an MCMC is to produce a series of samples in parameter
space (a Markov Chain) with the property that the density of
points is proportional to the probability distribution (the target
density) one is interested in mapping, in our case the posterior
pdf of Eq.~\eqref{eq:bayes}. There are several algorithms that can
produce a chain with the required properties. Here we employ the
Metropolis--Hastings algorithm
\cite{Metropolis:1953am,Hastings1970}: the chain is started from a
random point in parameter space, $\params_0$, and a new point
$\params_1$ is proposed with an arbitrarily proposal density
distribution $q(\params_n,
\params_{n+1})$. The transition kernel
$T(\params_n,\params_{n+1})$ gives the conditional probability for
the chain to move from $\params_n$ to $\params_{n+1}$, and it must
satisfy the ``detailed balance'' condition
 \begin{equation}
\pdf(\params_{n+1} \vert \data)T(\params_{n+1},\params_{n}) =
\pdf(\params_{n} \vert \data)T(\params_n,\params_{n+1})
\end{equation}
 so that the posterior $\pdf(\params \vert \data)$ is the
stationary distribution of the chain. This is achieved by defining
the transition kernel as
 \begin{align}
  T(\params_n,\params_{n+1}) & \equiv q(\params_n, \params_{n+1})
 \alpha (\params_n, \params_{n+1}) ,\\
 \alpha (\params_n, \params_{n+1}) & \equiv
 \min \left\{1, \dfrac{\pdf(\params_{n+1} \vert \data) q(\params_{n+1}, \params_{n})}
 {\pdf(\params_{n} \vert \data) q(\params_{n}, \params_{n+1})}
 \right\} ,
 \end{align}
 where $\alpha (\params_n, \params_{n+1})$ gives the probability
that the new point is accepted. Since $\pdf(\params \vert \data)
\propto P(\data \vert \params) \pdf(\params)$ and for the usual
case of a symmetric proposal density, $q(\params_n,
\params_{n+1})= q(\params_{n+1}, \params_{n})$, the new step is
always accepted if it improves on the posterior, otherwise it is
accepted with probability $P(\data \vert \params_{n+1})
\pdf(\params_{n+1})/P(\data \vert \params_{n}) \pdf(\params_{n})$.
The result is a sample list from the target distribution, from
which all the statistical quantities of interest can readily be
evaluated. Further details about MCMC methods can be found e.g. in
\pcite{MKbook}.

Our Bayesian MCMC analysis allows us to not only to determine
efficiently the best--fit value of the parameters, but also to
explore correlations between the model parameters and estimate
marginalized high probability regions, to which we now turn our
attention.

\section{Results and discussion}
\label{sec:results}

As mentioned above, we investigate two different dust correction
schemes for SFR data at low ($z<3$) redshift, one termed ``normal
dust correction'' and the other ``high dust correction''. This is
expected to roughly bracket the range of possible corrections. The
outcome of our analysis is strongly dependent on which dust
correction one chooses to employ, with the normal dust correction
implying hierarchical star formation, while the high dust
correction favours the monolithic scenario.

\subsection{Best fit models and parameter constraints}

\begin{figure*}
 \centering
 \includegraphics[width=\half\linewidth]{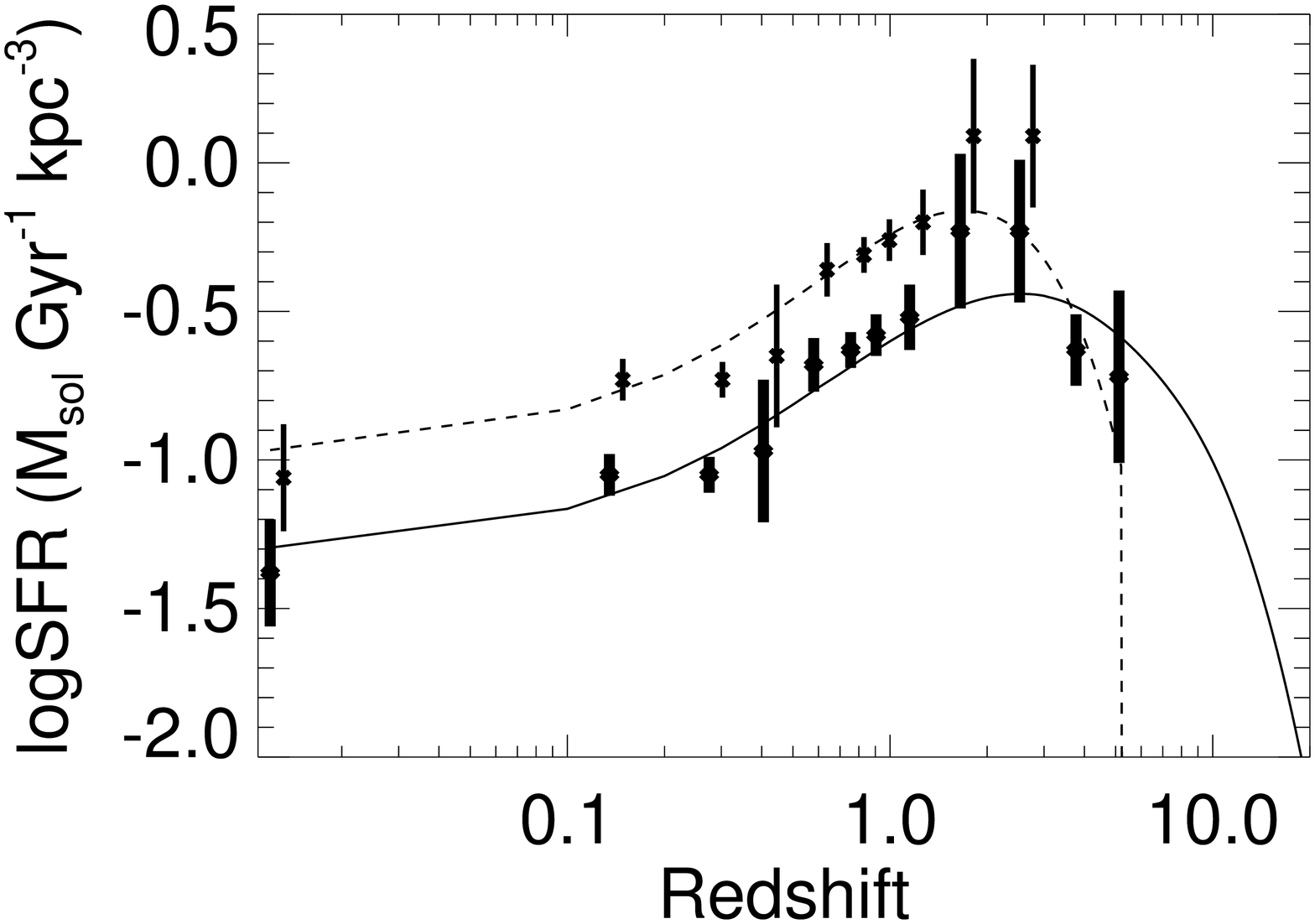}\hfill
 \includegraphics[width=\half\linewidth]{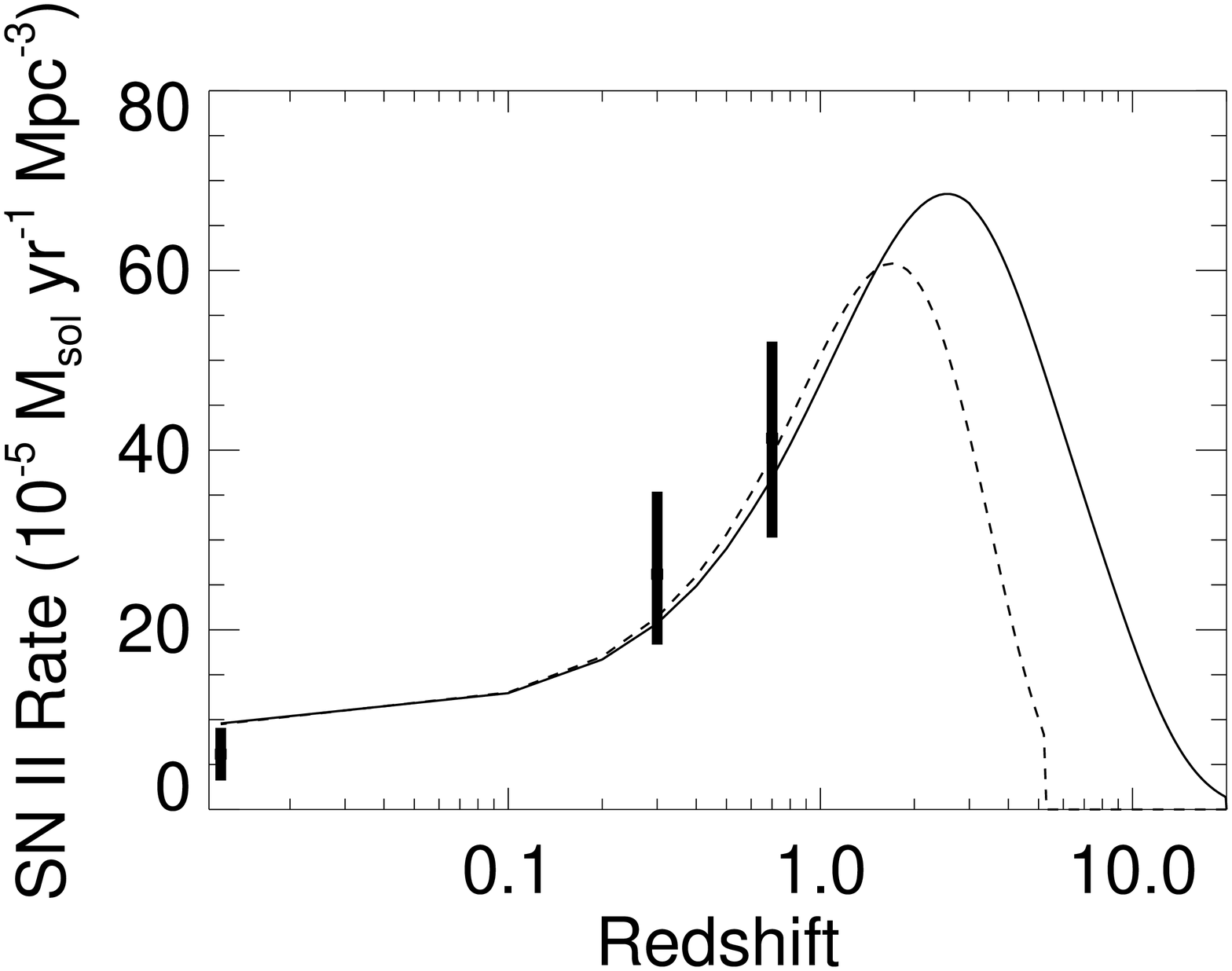}\\
 \includegraphics[width=\half\linewidth]{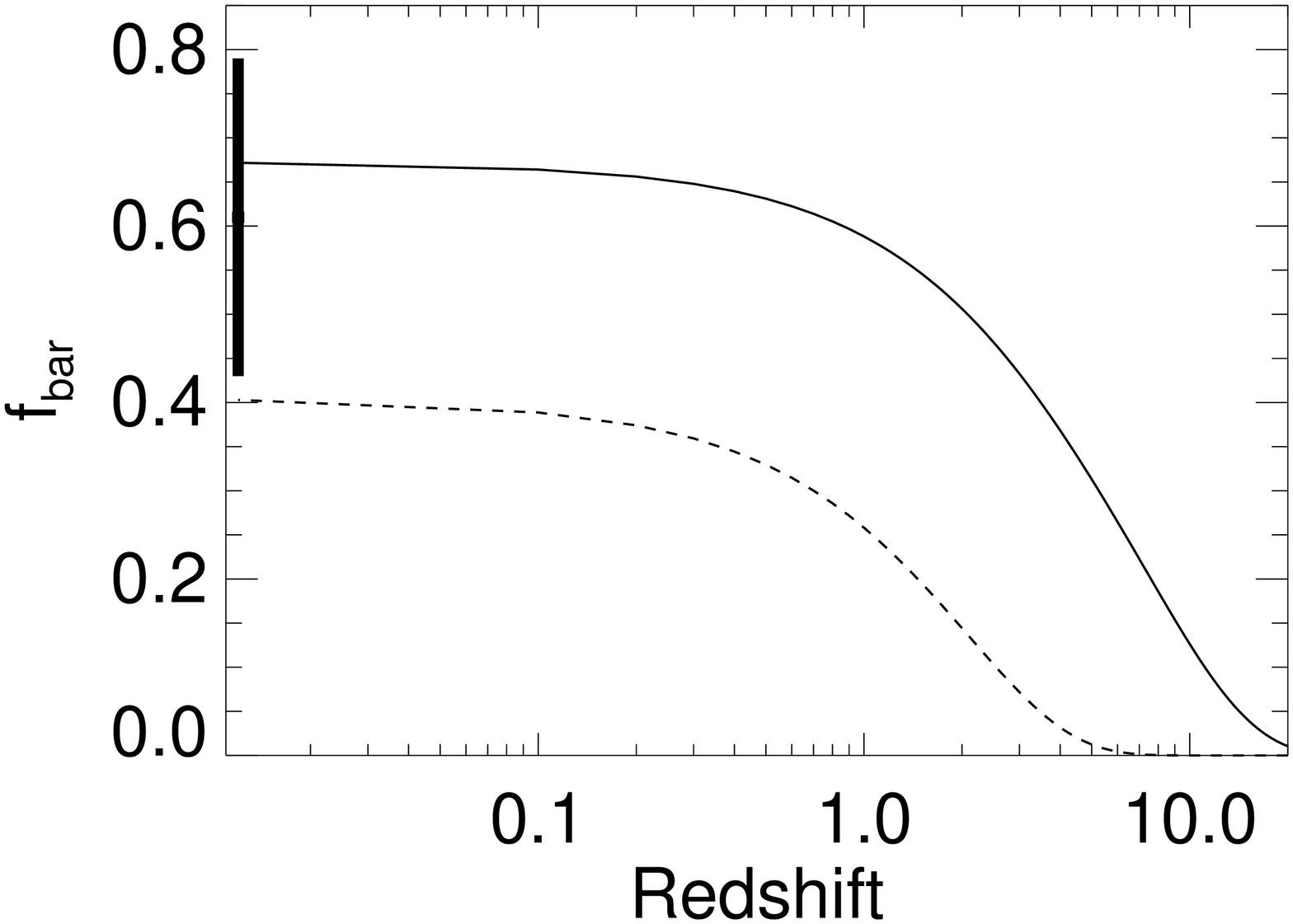}\hfill
  \includegraphics[width=\half\linewidth]{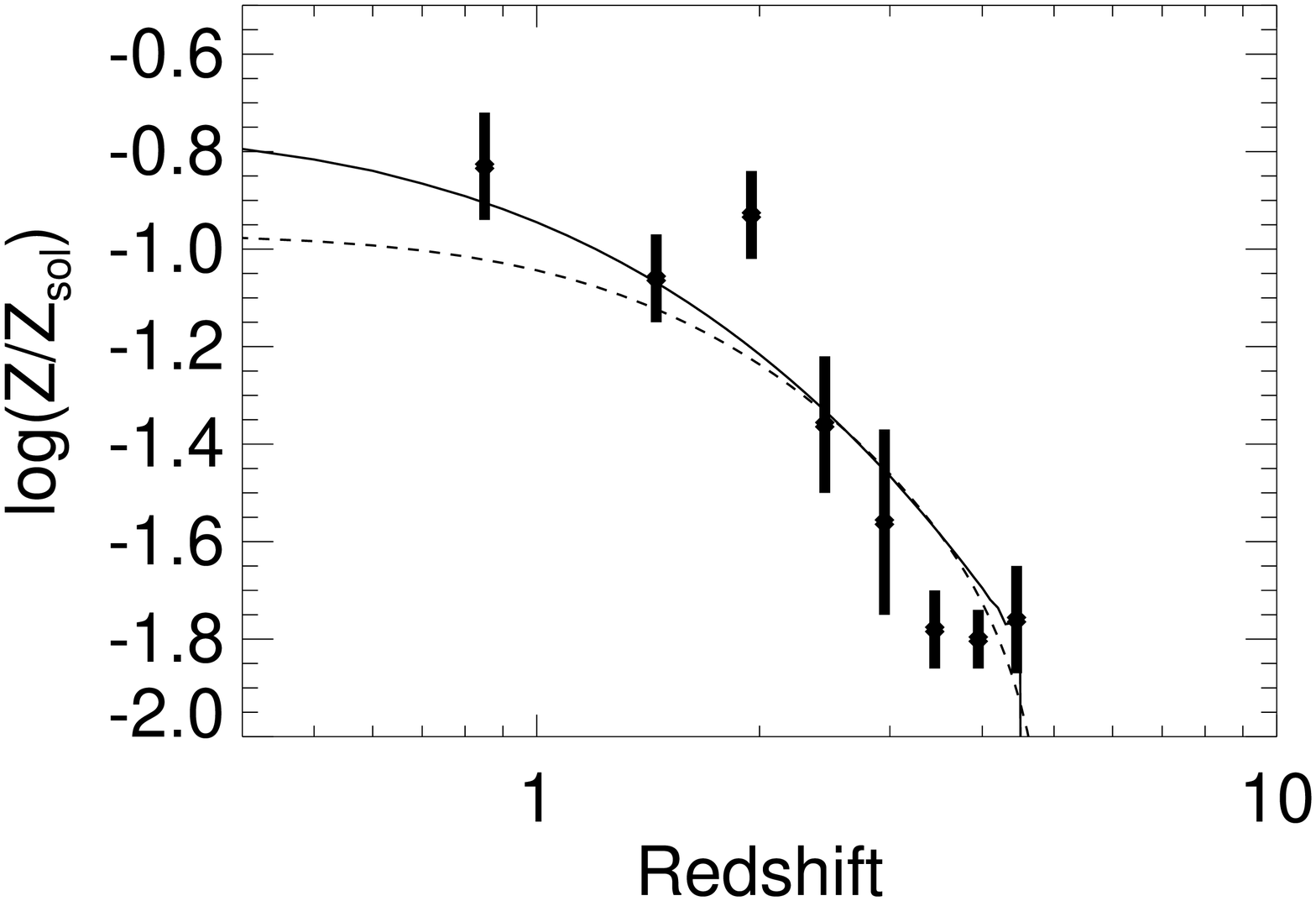}\\
 \caption{Best--fitting models for the
normal (solid line, hierarchical star formation, $\chi^2=26.60$)
and high (dashed, monolithic scenario, $\chi^2=33.3$) dust
corrections, with parameters as in Table~\ref{Table.Best_Fit}. In
the top left panel, showing the SFR, the low redshift ($z\leq3$)
data have been corrected for dust employing a normal dust
correction (1.0 mag, lower data points) or a high dust correction
(1.8 mag, upper data points). The high dust correction data have
bee shifted slightly to the right for display purposes.}
\label{fig:bestfit}
\end{figure*}

\begin{figure*}
 \centering
 \includegraphics[width=\half\linewidth]{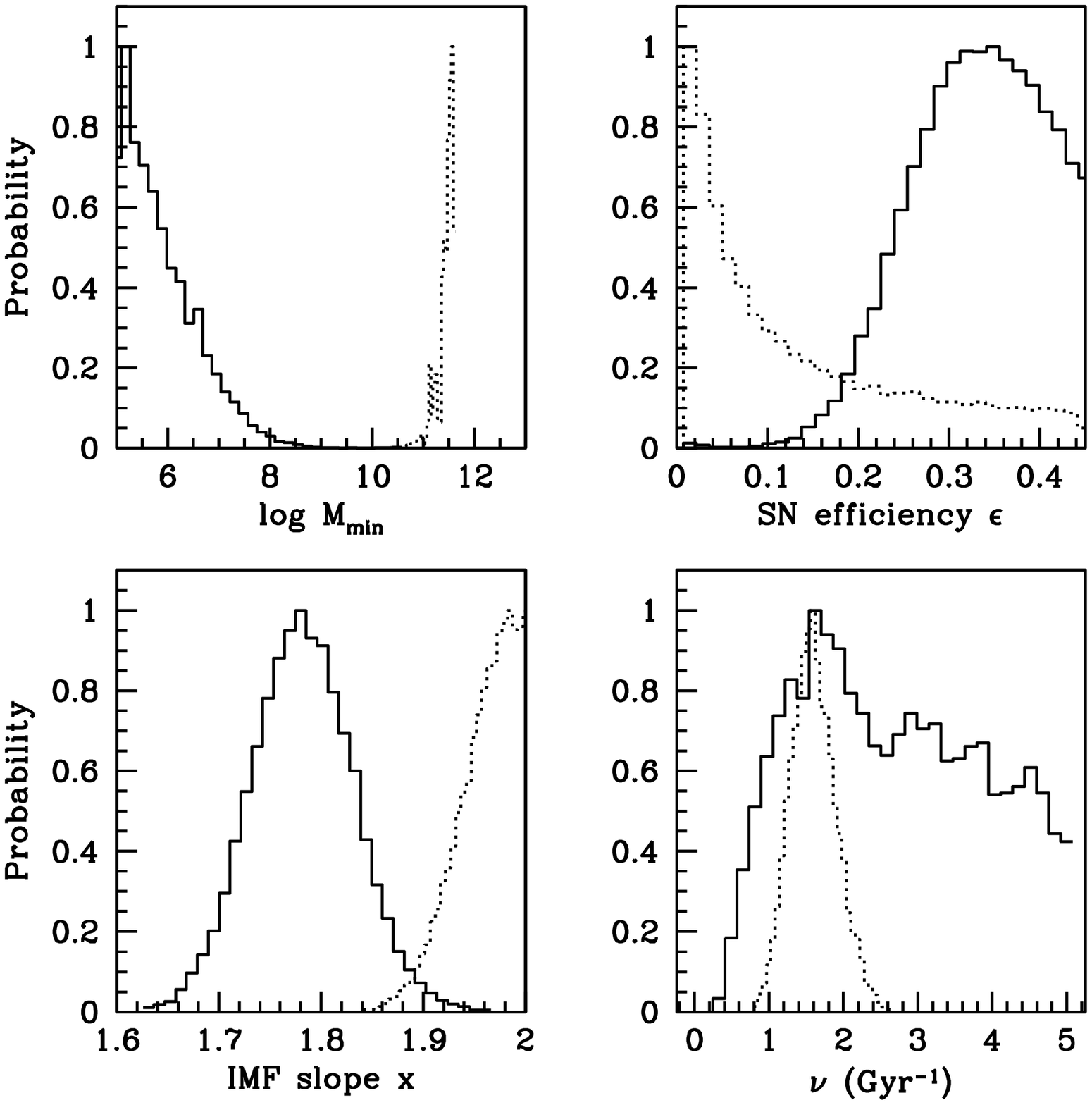}\hfill
 \includegraphics[width=\half\linewidth]{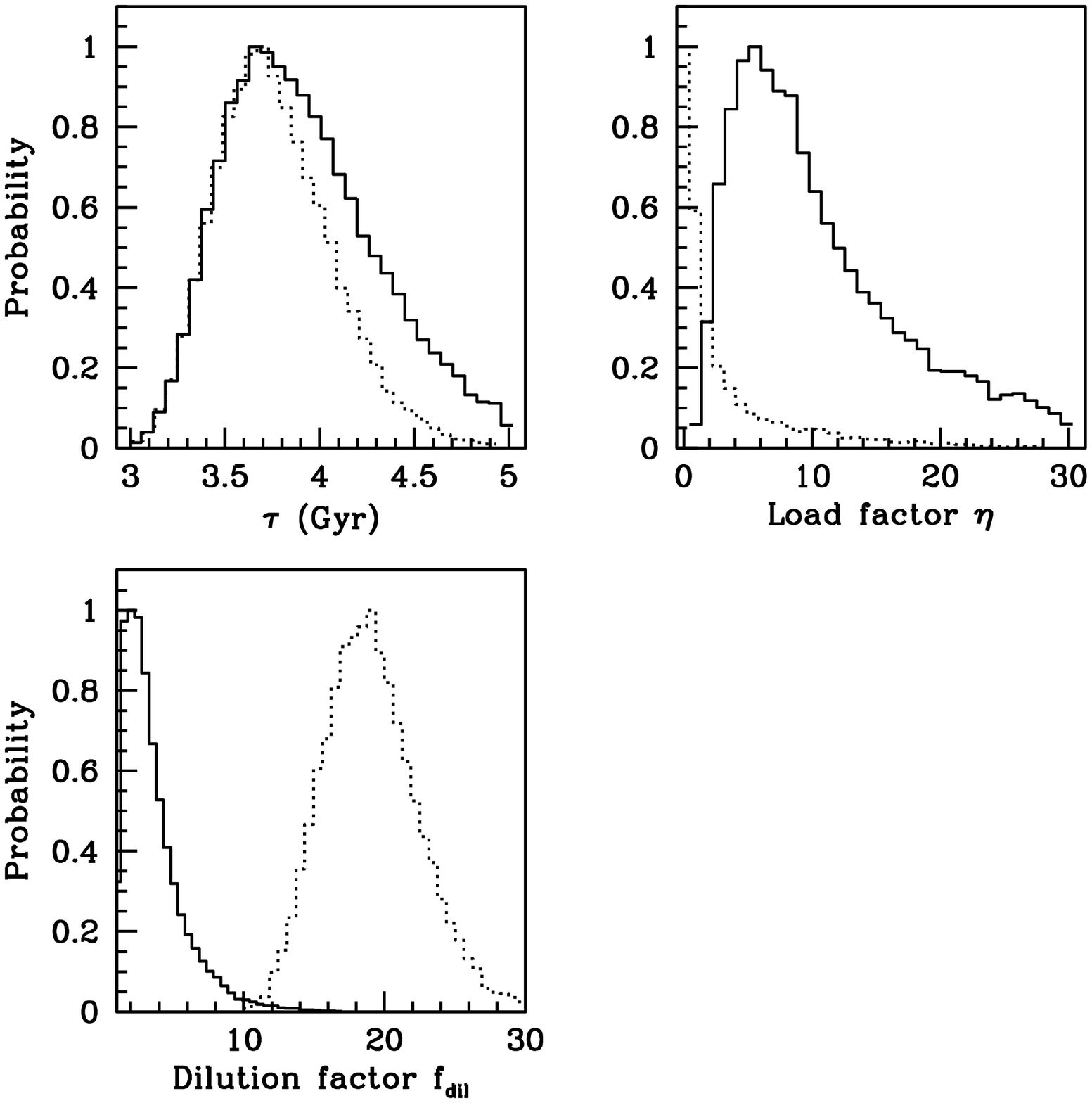}
\caption{1--dimensional marginalised posterior probability
distributions of the model parameters (normalized to their peak
values). Solid histograms are for the normal dust correction case
(hierarchical scenario), dotted for the high dust correction
(monolithic model).} \label{fig:1D}
\end{figure*}

The values of the best--fit model parameters for both dust
correction schemes are given in Table~\ref{Table.Best_Fit}, and
the corresponding SFR, SN rate, metallicity evolution and baryonic
fraction in structures are shown in Figure~\ref{fig:bestfit}. The
1--dimensional posterior probability distributions (with all other
parameters marginalized, i.e., integrated over) are plotted in
Figure~\ref{fig:1D}.

\begin{table*}
\begin{minipage}{177mm}
\centering
  \begin{tabular}{|l|r r r | r r r |}
    \hline
    Parameter & \multicolumn{3}{c}{Normal dust correction} & \multicolumn{3}{c}{High dust correction}\\
              & Best fit  & $68\%$ range & $95\%$ range &  Best fit  & $68\%$ range& $95\%$ range\\  \hline
  $\logM$     & 5.17  & $<6.11$ &  $<7.33$ & 11.60 & $(11.27,11.61)$ & $(10.97,11.61)$\\
  $\epsilon$  & 0.32  & $>0.30$ &  $>0.21$ & 0.23  & $<0.17$ & $0.39$\\
  $x$         & 1.77  & $(1.73,1.82)$ & $(1.68,1.87)$ & 1.97 & $>1.95$ & $>1.90$\\
  $\nu$ (Gyr$^{-1}$)& 4.15  & $>2.20$& $>1.01$ & 1.81 & $(1.29,1.90)$ & $(1.04,2.25)$\\
  $\tau$ (Gyr)& 3.59  & $(3.48,4.22)$ &  $(3.23,4.69)$ & 3.65 & $(3.45,4.07)$ & $(3.24,4.46)$\\
  $\eta$      & 8.74 & $(4.21,15.70)$ & $(2.24,24.49)$ & 0.02 & $< 3.80$ & $15.59$\\
  $\fdil$     & 2.72  & $(1.84,5.52)$ & $(1.27,9.34)$ & 20.75 & $(15.65,22.34)$ & $(13.06,26.28)$\\
    \hline
  $\chi^2$    & 26.60  & & & & 33.3 \\
  $\chi^2/\text{dof}$  & 1.6 & & & & 2.0\\
  \hline
  \end{tabular}
\caption{Best--fit parameter values and marginalized $68\%$ and
$95\%$ intervals for the normal (1.0 mag for $z\leq 3$) and high
($1.8$ mag for $z\leq3$) dust corrections. For cases where only an
upper or lower limit is found within our prior ranges, we give
one--tail intervals. We also give the best--fit chi--square and
the reduced chi--square, where the number of degrees of freedom
(dof) is 17 (for 24 data points and 7 free parameters).}
\label{Table.Best_Fit}
    \end{minipage}
    \end{table*}

We first discuss the case with the normal dust correction applied.
In order to fit the (dust-corrected) SFR at both high and small
redshifts, the model requires a small minimal mass ($\logM \sim
6$) and strong winds ($\epsilon \sim 0.3$). Although the value of
the supernova energy transfer parameter is quite large, it is not
too far away from theoretical predictions, which give an upper
limit of $\epsilon = 0.22$~\cite{larson74}. An IMF power--law
index $x \sim 1.8$, slightly larger then the Scalo IMF, is also
preferred, which translates in fewer available supernovae. This is
linked to the high value of $\epsilon$, since the energy transfer
is so efficient that a large number of supernovae is not needed to
get the appropriate feedback energy to reproduce the data sets.
The metallicity load factor $\eta$ can be connected with the IMF
power--law index $x$ and \pcite{dalcanton06} gives $\eta$ values
for a variety of IMFs. The value of $\eta$ for the \pcite{scalo86}
IMF ($x =1.7$, close to our best fit value, $x =1.77$) is $\eta =
16.8-18.6$, in reasonable agreement with our value, $\eta = 8.74$.
This leaves metal--rich outflows as the only viable mechanism for
producing the low effective yields observed in gas--rich galaxies.
in agreement with suggestions presented in \pcite{dalcanton06}.
The dilution factor $\fdil$ is of order 2, which again is very
reasonable, given the complex physics this parameter is supposed
to summarize. The value of the chi--square of the best fit model
in this case is $26.60$ for 17 degrees of freedom, which suggests
that our model captures the essential features of the data.
Figure~\ref{fig:bestfit} shows the best fit models for normal dust
correction~(solid line) and high dust correction~(dashed line).
Both models provide an acceptable fit to the data, although in the
normal dust correction case the low--redshift metallicity and the
present--day baryon fraction in structures appear in better
agreement with the data. For a redshift above $z\approx5$, the
metallicity of the hierarchical model drops very sharply to 0
because of the very significant winds.

Turning now to the high dust correction case, we notice that the
preferred values of the parameters in our model are very different
from the previous case. Most importantly, a high dust correction
at small redshift boosts the value of the SFR for $z\leq3$, and
this pushes our model to very large values of $\logM$, of the
order $\logM \sim 11-12$. This implies that star formation occurs
monolithically in heavy spheroids, as discussed in the
introduction. We expect dry mergers to play a significant role in
the build--up of massive  $\logM \sim 13$ ellipticals in agreement
with observations showing that present--day spheroidal galaxies on
average have undergone between 0.5 and 2 major dry mergers since
$z \sim 0.7$ (\pcite{bell06}). Furthermore, we see from
Figure~\ref{fig:bestfit} (dashed curves) that the onset of both
the SFR and metals build--up is significantly delayed in this
scenario, until about $z\approx5$. The supernovae energy transfer
parameter $\epsilon$ becomes essentially irrelevant for such large
values of the minimum mass, since the potential is deep enough to
retain the ejected gas. The peak in the probability distribution
for $\epsilon$ observed in Figure~\ref{fig:1D} is therefore mostly
a consequence of a volume effect of our Bayesian MCMC scanning
technique. The star formation timescale $\tau \sim 3.5$ Gyr is in
good agreement with theoretical models for Milky Way size disk
galaxies (with virial mass $\log M_{\rm vir}$ close to our best
fit value for $\logM$). The IMF index is tilted towards extreme
values, thus reducing the SN rates but boosting the SFR (compare
Figure~\ref{fig:Sen_X}). This in turns increases the metallicity,
and a large dilution factor, $\fdil \sim 20$, is required to bring
the predictions in line with observations. We notice that this
agrees within a factor of two with the value already found in
previous works on the metallicity of SN ejecta, which was of order
10. However, the extremely steep IMF that this model prefers ($x
\approx 2$) appears to exclude the possibility that stellar
explosions are the main mechanism that drive galactic winds. This
is reasonable, since supernova-driven gas flows cannot escape from
massive galaxies' potential wells. A resolution to this wind
dilemma could come from the hypothesis of  supermassive black
hole~(SMBH) induced outflows~\cite{silk05}. In fact, the very low
value of the load factor~($\eta = 0.02$) is consistent with this
scenario since the SMBH undergoes most of its growth in the gas
rich phase and its outflow expels mostly unprocessed gas. Although
our model does not include the physics of SMBHs, it is tempting to
say that our best fit model suggests that SMBHs should play a key
role in the evolution of massive spheroids.

In general, we observe that the high dust correction case seems to
stretch our model parameters to extreme values, suggesting either
a strong tension between datasets (mostly SFR and metallicity
data) or a failure of the model to fully encapsulate all of the
relevant physical processes. Even though with a reduced
chi--square per dof of $2.0$ this scenario is less favoured than
the hierarchical star formation model discussed above, it appears
that the monolothic formation model cannot be dismissed yet. It is
interesting that our 7 parameter model is able to describe both
cases, and that the SFR dust correction plays a major role in
defining which scenario is preferred.

\subsection{Correlations among parameters}

\newcommand{\quarter}{0.24}
\begin{figure*}
 \centering
 \includegraphics[width=\quarter\linewidth]{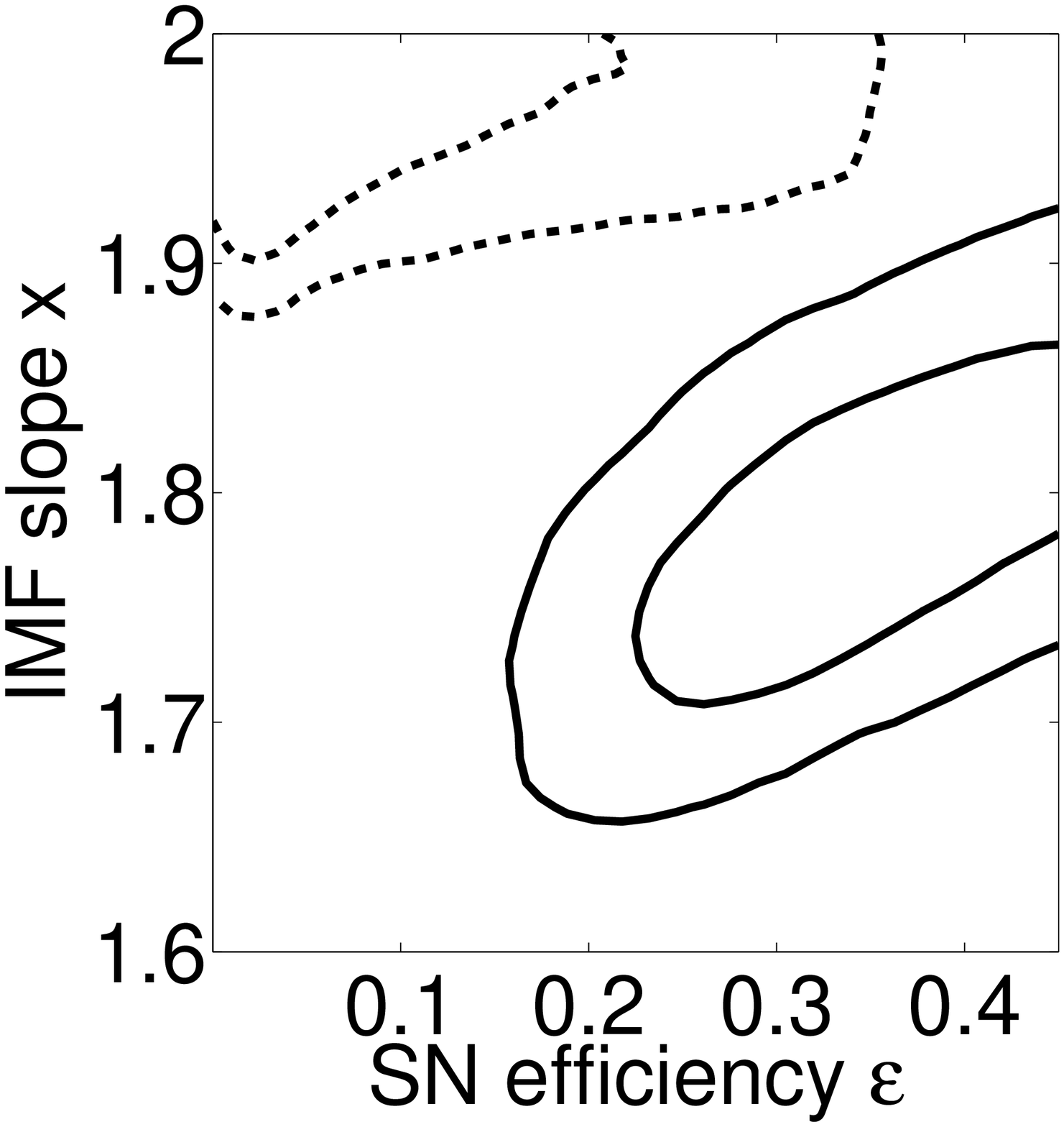} \hfill
 \includegraphics[width=\quarter\linewidth]{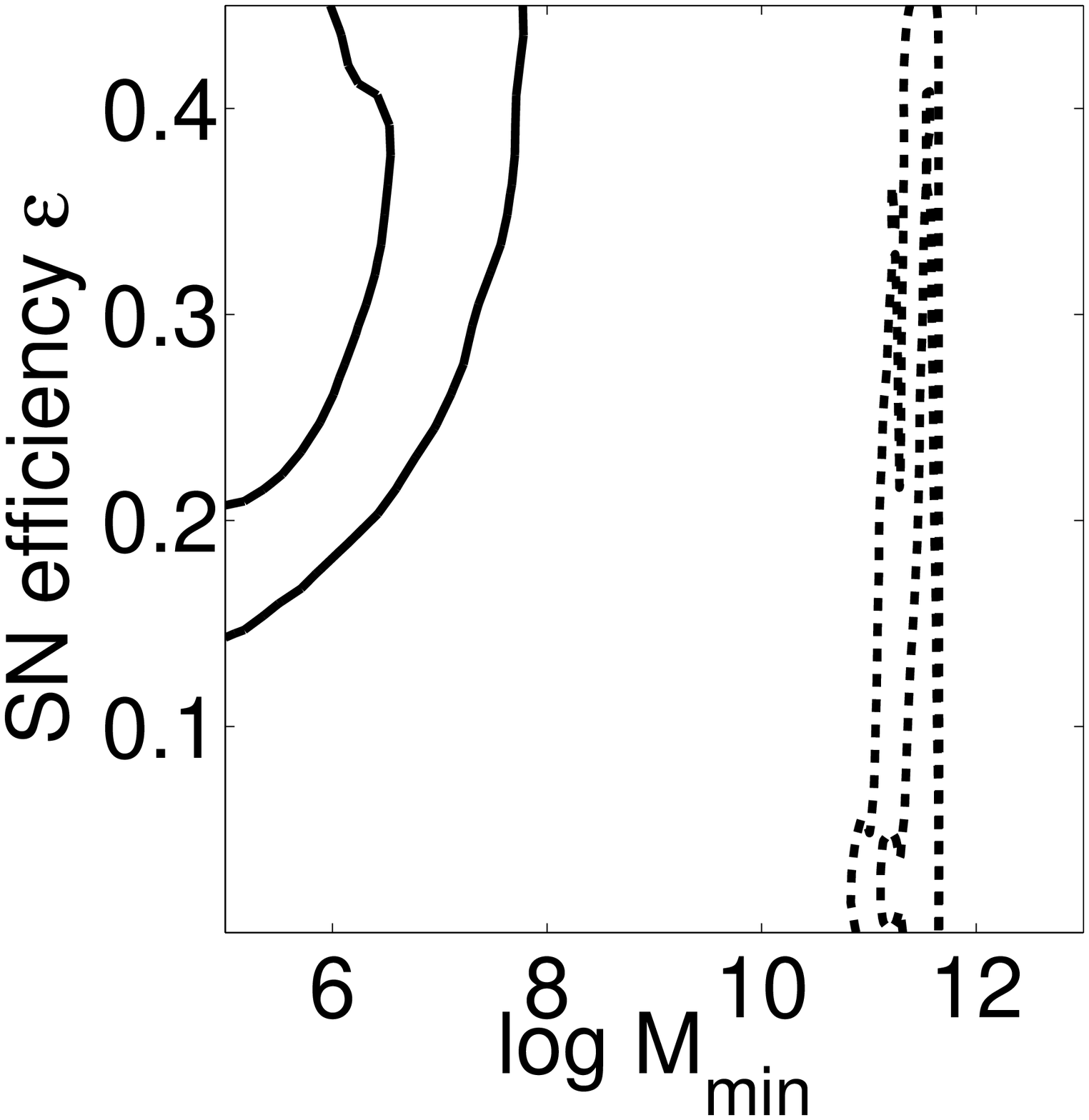} \hfill
 \includegraphics[width=\quarter\linewidth]{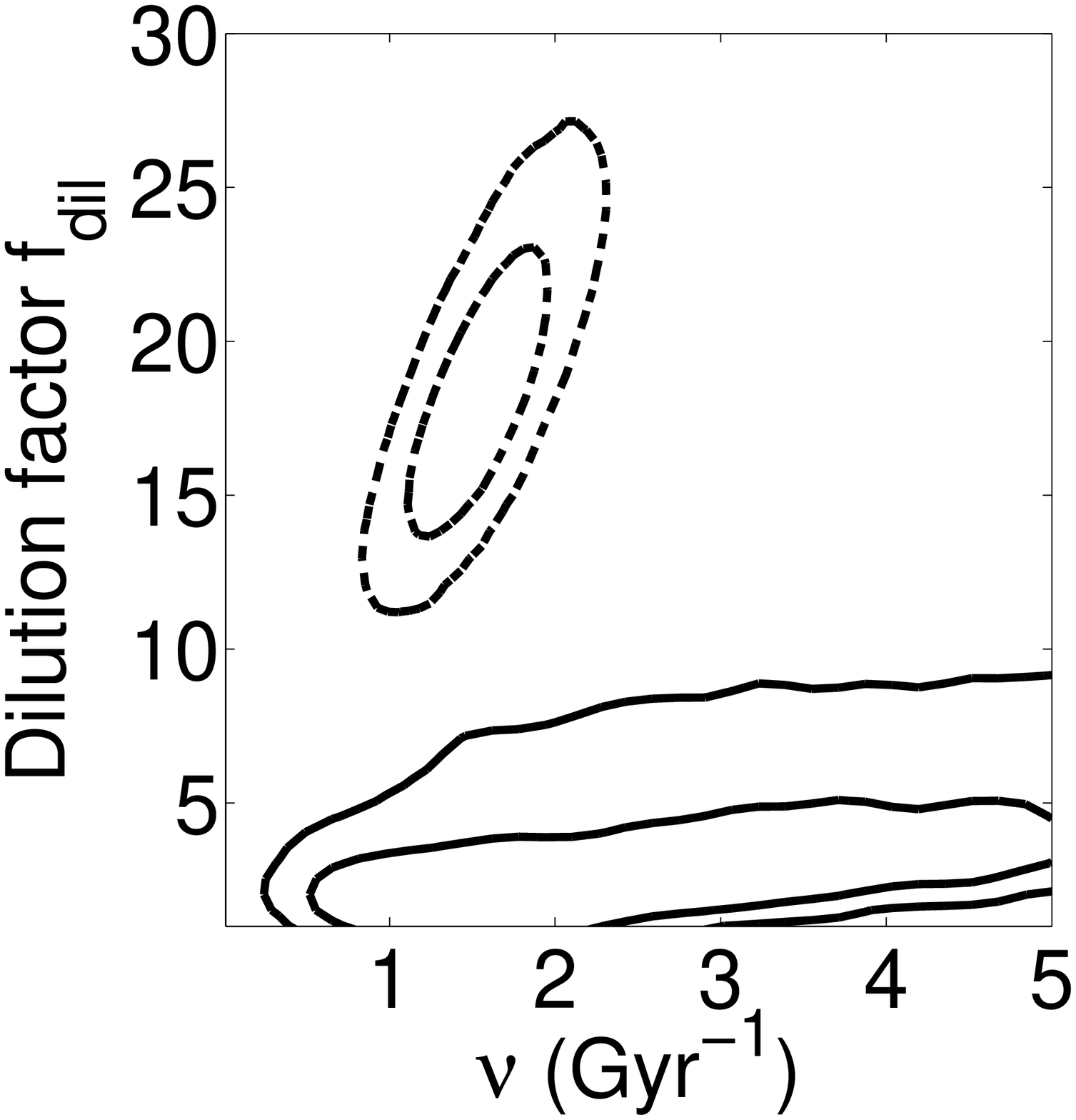} \hfill
 \includegraphics[width=\quarter\linewidth]{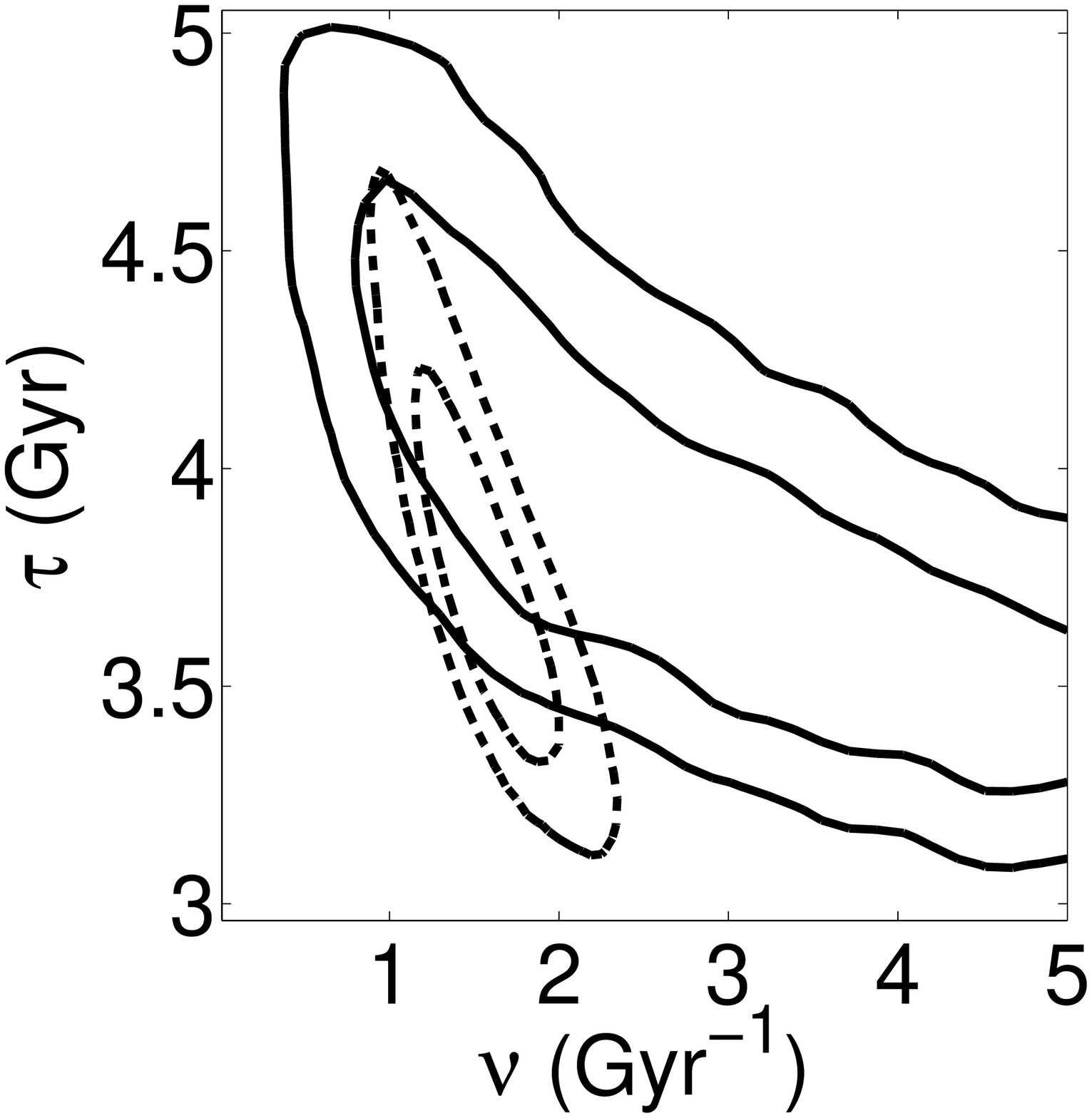}
 \caption{Contours enclosing joint 2D $68\%$ and
$95\%$ regions, with all other parameters marginalized, for both
the ``high dust correction'' (dashed, monolithic scenario) and the
``normal dust correction'' case~(solid, hierarchical star
formation).} \label{fig:2D}
\end{figure*}

We now turn to discuss the most relevant correlations among the
model parameters in light of their physical interpretation and of
their impact on the observables, as shown in
section~\ref{sec:Sensitivity}. Figure~\ref{fig:2D} shows a
selection of 2D joint posterior probability distributions for
$\logM$, ${\epsilon}$, $x$, $\nu$, $\tau$, and ${\fdil}$, thus
giving complementary information to the 1D distributions plotted
in Figure~\ref{fig:1D}. The contours enclose joint 2D $68\%$ and
$95\%$ regions, with all other parameters marginalized, for both
the ``high dust correction'' (dashed) and the ``normal dust
correction'' case~(solid).

In the first panel on the left of Figure~\ref{fig:2D}, showing the
$x-\epsilon$ plane, we observe a positive correlation between the
IMF power--law index and the SN type II energy efficiency factor.
This is expected, since an IMF with a higher power--law index
produces less SNe, each of which has to contribute more energy,
leading to higher values for the parameter ${\epsilon}$ (compare
Figures~\ref{fig:Sen_E} and \ref{fig:Sen_X}, panels showing the
SFR and metallicity dependence). For the``high dust correction''
case (dashed lines) this correlation is weaker, confirming our
conclusion that SNe type II cannot drive the winds in the massive
spheroids. The $\epsilon-\logM$ plane shows that for structures of
smaller mass (``normal dust correction'' case, solid lines) the
parameter $\epsilon$ needs to be large, while for high mass
structures (as preferred in the``high dust correction'' case,
dashed curves), $\epsilon$ is essentially unconstrained,
indicating that SNe feedback is irrelevant for massive spheroids.
The different physical processes taking place in small structures
and massive spheroids can be further investigated by looking at
the correlations in the $\fdil-\nu$ plane. We expect to find a
positive correlation among $\nu$ and the dilution factor $\fdil$,
as larger $\nu$ increases the SFR (compare Figure~\ref{fig:Sen_V})
thus leading to a more metal--rich ISM. To bring this back in line
with the data, a larger dilution factor is needed. The above line
of reasoning explains the strong positive correlation one observes
for the high mass structures (dashed) where winds do not play a
strong role and metals cannot escape from the structure. In
contrast, metal--rich winds are dominant for smaller structures
(solid curves), thus expelling most of the metals produced. This
results in almost no correlation between ${\fdil}$ and $\nu$,
since the impact of $\nu$ on the SFR and metallicity predictions
can be mimicked by a different combination of values for
$\epsilon$ and $\logM$. Finally, in the right--most panel of
Figure~\ref{fig:2D}, we display the probability distribution in
the $\tau-\nu$ plane, which exhibits a strong negative
correlation. Again, this is expected on the grounds that large
values of the parameter $\nu$ increase the SFR (compare
Figure~\ref{fig:Sen_V}) and a smaller time--scale is thus required
in order to quench star formation fast enough (see
Figure~\ref{fig:Sen_T}).

\section{Conclusions}
\label{sec:conc}

We have presented a well--motivated physical model of the cosmic
star formation incorporating supernova feedback, gas accretion and
enriched outflows. We computed the cosmic star formation history
and the chemical evolution in the interstellar medium of forming
galaxies as a function of redshift, and we presented for the first
time a full statistical treatment of the observational data, which
accounts for the possibility of systematic errors in the data
sets.

We have employed four different observational data --- the
observed cosmic star formation rate up to $\rm z \sim 5$, the
observed rate of type II supernova up to $\rm z \sim 0.7$, the
present baryon fraction in structures and the evolution of the
metal content in the ISM --- to derive constraints on the free
parameters of our model. After employing a Bayesian procedure to
rebin the SFR and SN rate data, we found that the low redshift
($z\leq3$) SFR dust correction adopted has a critical impact on
the scenario favoured by the data.

For what we have termed ``normal dust correction'', the
hierarchical star formation model is preferred, where star
formation occurs in small structures first and supernovae winds
are important. While the winds load factor remains poorly
constrained, we can conclude that larger values are preferred, in
agreement with previous work \cite{dalcanton06}. Applying a larger
dust correction at small redshifts, we found that the data on the
contrary favour high values for the minimum mass of a dark halo of
the collapsed structures (monolithic star formation scenario).
This case requires a large dilution factor, a rather extreme IMF
slope and a fairly small winds load factor, at the model
parameters are pushed at the boundaries of the available range. We
have suggested that this might be interpreted in terms of the
presence of outflow from supermassive black holes, but this
possibility will require further investigation. It is worth
noticing that the monolithic star formation scenario has very
little star formation beyond $z \sim 5$. Observations of the
E--mode polarization power spectrum of the cosmic microwave
background however indicate that the Universe was re--ionized
around $z \sim 11$ \cite{spergel06}. This means that in this
scenario the reionization mechanism has to be found elsewhere than
in massive UV---emitting stars. Several alternatives have been
explored in the literature, for example reionization by decaying
particles (\pcite{hansen04}), or a high--redshift population of
mini--quasars that can reionize the IGM up to $50\%$ ionisation
fraction (\pcite{dijkstra04}).

For both models, the IMF slope is  large. Unfortunately, this does
not help in distinguishing one model from the other, since
observations have so far not yielded convincing results concerning
the form of the stellar IMF or its variations in space and
time~\cite{scalo98}. The most important difference among the IMFs
is that the fraction of high--mass stars is larger for a shallower
IMF. Since only high--mass stars emit significant amount of
ultraviolet light, this results in a spectrum which is more
shifted towards the ultraviolet for a typical galaxy with an e.g.
Salpeter IMF~($x=1.35$) as compared with a Scalo IMF~($x=1.7$). In
turn, this leads to a different reionization history, which can be
in principle compared with the optical depth to reionization as
inferred from cosmic microwave background polarization
measurements.

While the monolithic scenario is less preferred in terms of
quality of fit, it is clear that more work is required to be able
to draw firm conclusions as to the viability of the two different
models. Of particular importance remains the statistical treatment
of the data, for which we have here presented a new procedure that
we hope will prove useful for future work.

\section*{Acknowledgments}

The authors would like to thank an anonymous referee for several
interesting comments. RT is supported by the Royal Astronomical
Society through the Sir Norman Lockyer Fellowship, and by St
Anne's College, Oxford.

\appendix

\newcommand{\vv}{\mathcal{V}}
\newcommand{\vz}{\mathcal{Z}}

\section{Binning of data accounting for undetected systematics}
\label{app}

\subsection{No redshift uncertainty}
\label{app_noz}

\begin{figure}
\centering
\includegraphics[width=\linewidth]{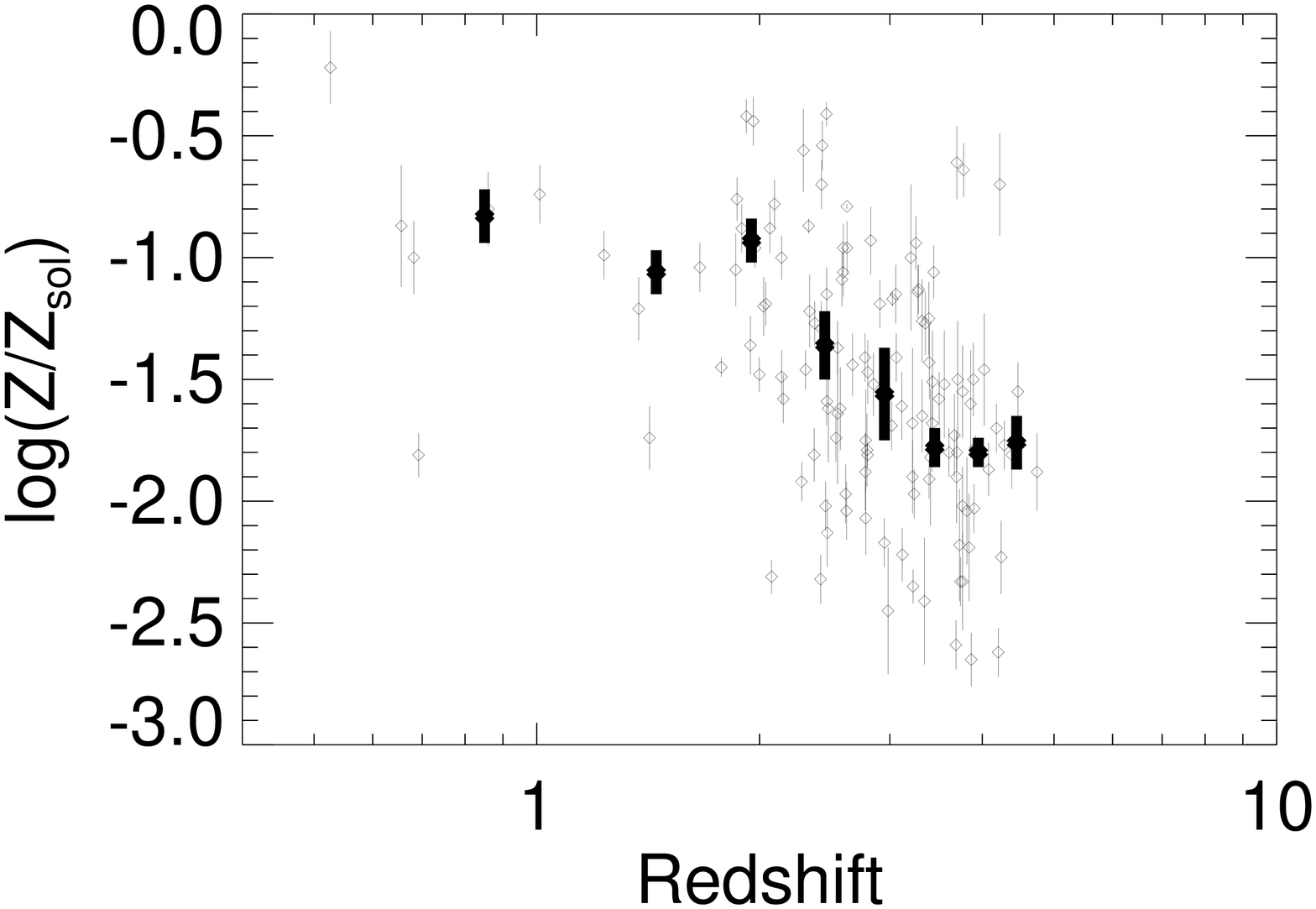}
\caption{Raw metallicity data and the binned values after the
statistical treatment.} \label{fig:metals}
\end{figure}

We wish to  define $B$ bins in redshift space. Within each bin
$b$, $1\leq b \leq B$, we have a collection of measurements (in
our case, metallicity or SFR data), each with its own statistical
accuracy and possibly an unspecified systematic error. That
systematic differences above the quoted statistical errors
dominate the raw data is apparent from a plot of the unbinned
metallicity or SFR observations, that show a scatter of up to an
order of magnitude for observations at about the same redshift.
The origin of the systematic discrepancy can vary, from
underestimated statistical errors in the observation to intrinsic
dispersion in the observed systems to differences in the way the
data are collected. In the presence of systematic errors, we
cannot simply take the weighted average of the data within each
bin. Instead, we model the presence of unknown systematics as
follows.

Let us consider the measurement of a quantity $y_b$ in a top--hat
bin $b$, $1\leq b \leq B$ --- in our case, this represents the
metallicity value at the redshift of the bin, $z_b$, and we assume
we can neglect the redshift uncertainty of the measurements (this
issue is addressed in the next section). Each measurement consists
of a central value $d_i$ and a statistical error $\sigma_i$,
$1\leq i \leq N_b$, for $N_b$ different measurements within bin
$b$. If the $i$--th datum does not suffer from a systematic error
(or where the systematic error, $S_i$, is negligible compared with
the quoted statistical error), the likelihood function is modelled
as a Gaussian with the quoted standard deviation $\sigma_i$:
\begin{equation}
P_{i,g}(d_i|y_b) = \frac{1}{\sqrt{2\pi}\sigma_i}
\exp\left[-\frac{1}{2}\left(\frac{d_i-y_b}{\sigma_i}\right)^2\right].
\label{eq:good_points}
\end{equation}
For the sake of brevity, let us denote such measurements as
``good'' measurements, as indicated by the subscript $g$. If the
datum suffers from an undetected systematic, i.e. the dominant
error is $S_i \gg \sigma_i$, the likelihood is instead given by
(neglecting the statistical error wrt the systematic one)
\begin{equation}
P_{i,s}(d_i|y_b) = \frac{1}{\sqrt{2\pi}S_i}
\exp\left[-\frac{1}{2}\left(\frac{d_i-y_b}{S_i}\right)^2\right],
\label{eq:bad_points}
\end{equation}
where the subscript $s$ denotes ``systematics'', or ``spoiled''
measurements, for brevity. Now of course we do not know which
measurements suffer from systematic, but this can be determined
statistically using the following procedure (adapted
from~\pcite{press96}).

We denote by $p$ the probability that each of the measurements $i$
in bin $b$ is a ``good'' one. Conversely, $1-p$ is the probability
that the datum suffers from systematics. Furthermore, we include a
binary vector $\vv=(\vv_1,\dots, \vv_{N_b})$, whose elements
$\vv_i$ ($1\leq i\leq N_b$) can either be 0 or 1, determining
whether the datum $i$ is a good one (for $\vv_i=1$) or a spoiled
one (for $\vv_i=0$). We can then compute the posterior probability
for the value of the observed quantity $y_b$ in bin $b$ by
multiplying the individual contributions of the measurements in
the same bin and marginalizing over the unknown quantities $p$ and
$\vv$ (see Eq.~(16) in \pcite{press96}):
\begin{equation}
P(y_b|d_b) \propto {\int}dp \prod_{i=1}^{N_b}\left[p P_{
g,i}+(1-p)P_{s,i}\right], \label{eq:prob_bin}
\end{equation}
where $d_b$ denotes the collection of measurements in bin $d$,
i.e. $d_b=(d_1,\dots,d_{N_b})$. For the prior probability on $p$
we have assumed a flat prior distribution between $0\leq p\leq 1$
and the proportionality factor might be determined by requiring
that the likelihood be normalized to unity, but this is not
necessary in our application. The precise numerical value of the
error associated with systematics, $S_i$ does not matter, as long
as $S_i \gg \sigma_i$. In our case, we take $S_i$ to be unity on a
log scale, corresponding to one order of magnitude uncertainty on
the observable.

From the posterior distribution \eqref{eq:prob_bin}, the central
value of the bin $b$ is obtained as the peak of the distribution,
while the standard deviation is defined as the range enclosing
$68.4\%$ ($1\sigma$ range) of the probability. These values are
given in Table~\ref{Table.Cosmic_Metal} for the metallicity data,
and are then used for the likelihood function employed in the fit
of the model. Of course one could as well employ the full
probability distribution of Eq.~\eqref{eq:prob_bin} as the
likelihood function, but for simplicity we have summarized it as a
Gaussian with mean and standard deviation computed as described
above. The collection of raw, unbinned metallicity data and the
resulting bins form our statistical treatment are shown in
Figure~\ref{fig:metals}.

\subsection{Accounting for redshift uncertainty}
\label{app_z}
\begin{figure}
\centering
\includegraphics[width=\linewidth]{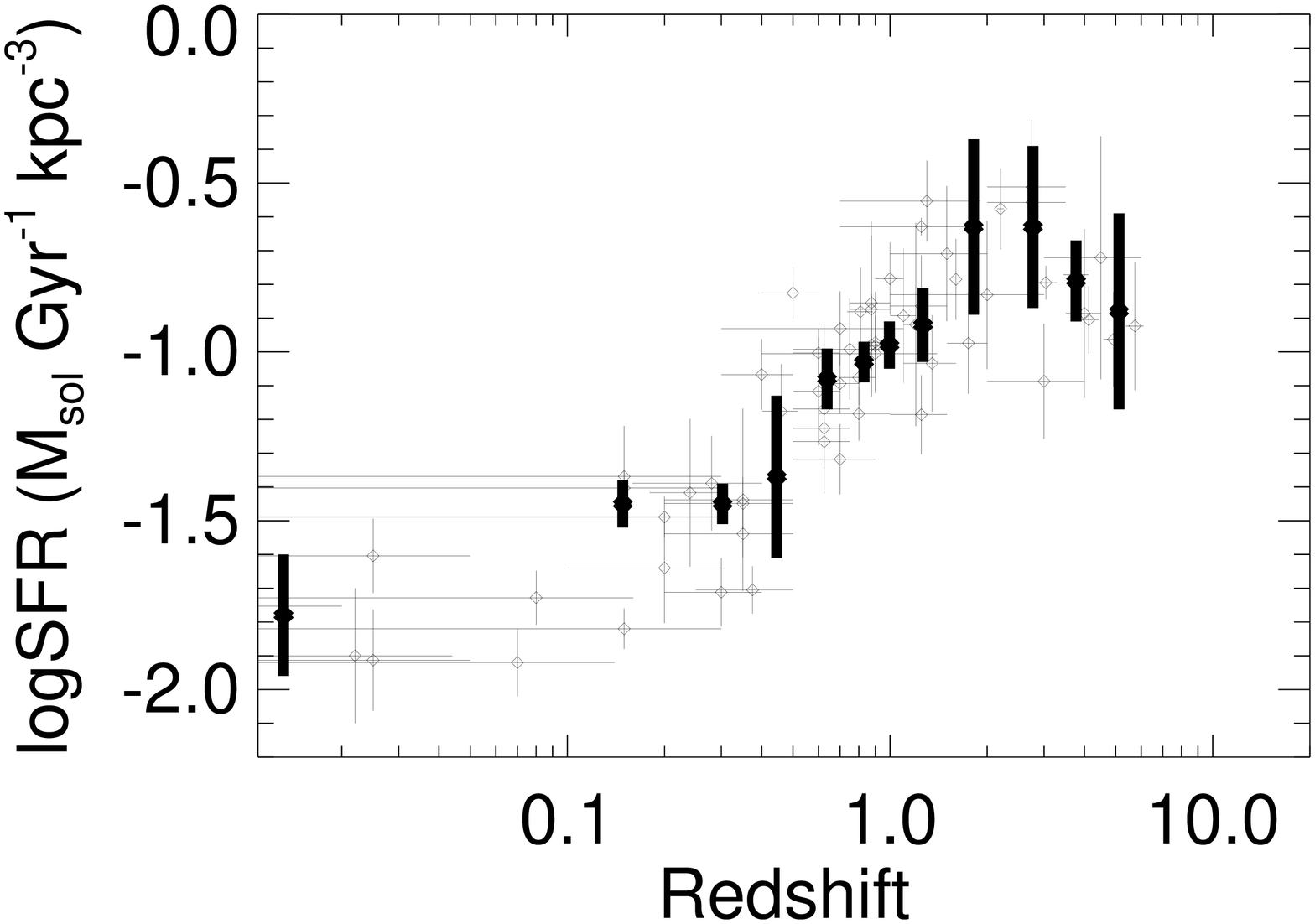}
\caption{Raw SFR data and the binned values after the statistical
treatment including redshift uncertainties. No dust correction has
been applied to the data at this stage.} \label{fig:SFRbin}
\end{figure}

When the observed quantity suffers from a substantial redshift
uncertainty, as in the case of the SFR data, we need to take into
account the redshift error in our binning procedure, as this
introduces a further uncertainty as to which bin a given datum
belongs to. The above procedure is then modified as follows.

The probability that an observations with central redshift $z_i$
and redshift uncertainty $\tau_{i}$ belongs to the $b$--th
redshift bin (centered at redshift $z_b$) is modelled as a
Gaussian, i.e.
\begin{equation}
P(z_i|z_b) = \frac{1}{\sqrt{2\pi}\tau_{i}}
\exp\left[-\frac{1}{2}\left(\frac{z_i-z_b}{\tau_{i}}\right)^2\right].
\label{eq:redshift_points}
\end{equation}
Given the uncertainty on the location of the measurements in
redshift, it is now impossible to assign data points to top--hat
bins. Instead, one needs to marginalize over all possible
assignments of data points among redshift bins, with each point's
contribution weighted by the conditional probability of
Eq.~\eqref{eq:redshift_points}. For each bin, let us introduce a
new binary variable, $\vz=(\vz_i,\dots,\vz_N)$, whose elements
indicate whether the $i$--th datum ($1\leq i \leq N$) belongs to
the bin under consideration ($\vz_i=1$) or not ($\vz_i=0$). If we
knew which datum belongs to which redshift bin, then we could
assign an exact binary sequence to $\vz$ (this corresponds to the
case considered in the previous section). Instead, we sum
(marginalize) over all possibilities, writing for the posterior
probability of the SFR value $y_b$ at redshift $z_b$, given $d$,
the full collection of data points at all redshifts
 \begin{equation}
 P(y_b|d) = \sum_{\vz} P(y_b, \vz|d) = \sum_{\vz}
 P(y_b|\vz,d)P(z_i|z_b,\vz_i=1),
 \end{equation}
where the conditional probability $P(y_b|\vz,d)$ is given by
Eq.~\eqref{eq:prob_bin}, given a specific assignment for $\vz$.
The sum over $\vz$ can be replaced by a product of binomial terms,
so that we finally obtain, using Eqs.~\eqref{eq:prob_bin} and
\eqref{eq:redshift_points}
 \begin{equation}
 P(y_b|d) \propto \int dp~\prod_{i=1}^{N}\left\{\left[p P_{
g,i}+(1-p)P_{s,i}\right]P(z_i|z_b)\right\} -1.
\label{eq:prob_bin_z}
\end{equation}
Notice that the product is here over all points in the dataset,
not just over the ones in a bin, as in Eq.~\eqref{eq:prob_bin}.

Since we include the full dataset for each bin, the resulting
errors are in principle correlated across bins. However, the
Gaussian term of Eq.~\eqref{eq:redshift_points} ensures that only
``nearby'' points give a non--negligible contribution to the value
of bin $b$. We therefore consider it acceptable to ignore the
correlation among bins when using the mean and standard deviation
of Eq.~\eqref{eq:prob_bin_z} for the likelihood function for the
SFR. The results from this procedure are tabulated in
Table~\ref{Table.Cosmic_Sfr}, and are plotted alongside with the
raw, unbinned data in Figure~\ref{fig:SFRbin}.


\begin{thebibliography}{}

\bibitem[Bell et al.<(2006)>]{bell06}Bell~E.~F. et al., 2006, ApJ, 640, 241
\bibitem[Birnboim et al. <(2007)>]{birnboim07}Birnboim~Y., Dekel~A., Neistein~E., 2007, astro-ph/0703435
\bibitem[Bond~\&Efstathiou <(1999)>]{bond84}Bond~J.~R., Efstathiou~G., 1984, ApJ, 285, L45
\bibitem[Bower et al. <(2006)>]{bower06}Bower~R. et al. 2006, MNRAS, 370, 645
\bibitem[Cappellaro et al. <(1999)>]{cappellaro99}Cappellaro~E., Evans~R., Turatto~M., 1999, A\&A, 351, 459
\bibitem[Coles~\&~Lucchin <(1995)>]{coles95}Coles~P., Lucchin~F., 1995, Cosmology, John Wiley \& Sons Ltd
\bibitem[Condon <(1989)>]{condon89}Condon~J.~J., ApJ, 338, 13
\bibitem[Condon et al. <(2002)>]{condon02}Condon~J.~J., Cotton~W.~D., Broderick~J.~J., 2002, AJ, 124, 675
\bibitem[Croton et al <(2006)>]{croton06}Croton~D. et al, 2006, MNRAS 365,1
\bibitem[Daigne et al. <(2004)>]{daigne04}Daigne~F., Olive~K.~0., Vangioni--Flam~E,  Silk~J., Audouze, J.,
\bibitem[Daigne et al. <(2005)>]{daigne05}Daigne~F., Olive~K.~0., Silk~J., Stoehr~F., Vangioni--Flam~E., ApJ, 2006, 647, 773
\bibitem[Dalcanton <(2006)>]{dalcanton06}Dalcanton~J.,~J., preprint, astro-ph/0608590
\bibitem[Dahlen et al. <(2004)>]{dahlen04}Dahlen~T., et al. 2004, ApJ, 613, 189
\bibitem[Dijkstra et al. <(2004)>]{dijkstra04}Dijkstra~M., Haiman~Z., Loeb~A., 2004, ApJ, 613, 646
\bibitem[David et al. <(1990>]{david90}David~L.~P., Forman~W., \& Jones~C., 1990, ApJ, 359, 29
\bibitem[de Austri et al. <(2006)>]{de Austri:2006pe}  R.~R.~de Austri, R.~Trotta and L.~Roszkowski,  JHEP {0605} (2006) 002  
\bibitem[Efstathiou <(2000)>]{efstathiou00}Efstathiou~G., 2000, MNRAS, 317, 697
\bibitem[Forster-Schreiber et al. <(2006)>]{forster06}Forster-Schreiber~N. et al. 2006, ApJ, 645, 1062
\bibitem[Fukugita~\& Peebles <(2004)>]{fukugita04}Fukugita~M., Peebles~P.~J.~E. 2004, ApJ, 616, 643
\bibitem[Hammer et al. <(2005)>]{hammer05}Hammer~F. et al. 2005, A\&A, 430, 115
\bibitem[Hansen~\&~Haiman <(2004)>]{hansen04}Hansen~H.~H., Haiman~Z., 2004, ApJ, 600, 26
\bibitem[Hastings <(1970)>]{Hastings1970} Hastings W.K. 1970,{Biometrika}, 57, 97-109
\bibitem[Hopkins <(2004)>]{hopkins04}Hopkins~A.~M., 2004, ApJ, 615, 209
\bibitem[Jenkins et al. <(2001)>]{jenkins01}Jenkins~A., Frenk~C.~S., White~S.~D.~M., Colberg~J.~M., Cole~S., Evrard~A.~E., Couchman~H.~M.~P., Yoshida~N., 2001, MNRAS, 321 372
\bibitem[Larson <(1974)>]{larson74}Larson~R.~B., 1974, MNRAS, 169, 229
\bibitem[Lotz et al. <(2006)>]{lotz06}Lotz~J. et al. 2006,  astro-ph/0602088
\bibitem[MacKay <(2003)>]{MKbook} MacKay D., 2003, Information theory, inference, and learning algorithms. Cambridge University Press, Cambridge, U.K.
\bibitem[Maraston et al. <(2006)>]{mar06} Maraston c. et al. 2006, ApJ, 652, 85
\bibitem[Metropolis et al. <(1953)>]{Metropolis:1953am} Metropolis N., Rosenbluth A. W, Rosenbluth M. N., Teller A. H., and Teller, E. J. Chem. Phys., 21, 1087-1092
\bibitem[Pagel <(1997)>]{pagel97}Pagel~B.,~E. 1997, Nucleosynthesis and Chemical Evolution of Galaxies. Cambridge Univ. Press, Cambridge
\bibitem[Press <(1996)>]{press96}Press~W.~H., astro-ph/9604126
\bibitem[Press~\& Schechter <(1974)>]{press74}Press~W.~H., Schechter~P., 1974, ApJ, 187, 425
\bibitem[Prochaska et al. <(2003)>]{prochaska03}Prochaska~J.~X., Gawiser~E., Wolfe~A.~M., Castro~S., Djorgovski~S.~G., 2003, ApJ, 595, L9
\bibitem[Salpeter <(1955)>]{salpeter55}Salpeter~E.~E., 1955, ApJ, 121, 161
\bibitem[Scalo <(1986)>]{scalo86}Scalo~J.~M., 1986, Fundamentals of Cosmic Physics, 11, 1
\bibitem[Scalo <(1998)>]{scalo98}Scalo~J.~M., astro-ph/9811341
\bibitem[Schiminovich at al. <(2005)>]{schiminovich05}Schiminovich~D. et al. 2005, ApJ, 619, L47
\bibitem[Sheth~\& Tormen <(1999)>]{sheth99}Sheth~R.~K., Tormen~G., 1999, MNRAS, 308, 119
\bibitem[Silk <(2005)>]{silk05}Silk~J., 2005, MNRAS, 364, 1337
\bibitem[Spergel et al. <(2006)>]{spergel06}Spergel~D.~N., et al. 2006, astro-ph/0603449
\bibitem[Thomas et al. <(2005)>]{tho05} D. Thomas et al. 2005, ApJ, 621, 673
\bibitem[Trotta <(2007a)>]{Trotta:2005ar}  R.~Trotta, 2007a, Mon.\ Not.\ Roy.\ Astron.\ Soc.\  {378}, 72  
\bibitem[Trotta <(2007b)>]{Trotta:2007hy}  R.~Trotta,  2007b, Mon.\ Not.\ Roy.\ Astron.\ Soc.\  {378}, 819  
\bibitem[Worthey et al. <(1992)>]{worthey92}Worthey~G., Faber~S.~M.~\&~Gonzalez~J.~J., 1992, ApJ, 398, 69





\end{thebibliography}
\end{document}